\documentclass{aa}
\usepackage{txfonts}
\usepackage{graphicx,subfigure}
\usepackage{epsfig}
\usepackage{epsf}
\usepackage{lscape}
\usepackage{xcolor}
\usepackage{natbib}
\usepackage{rotating}
\usepackage{array}
\usepackage{varwidth}
\usepackage{supertabular}
\newcommand{\mic}{\,{\rm \mu m} }

\begin{document}

\title{Inferring the dust emission at submillimeter and millimeter wavelengths using neural networks}

\author{D. Paradis \inst{1} 
  \and
  C. Mény \inst{1}
  \and
  A. Noriega-Crespo \inst{2}
  \and
  K. Demyk \inst{1}
  \and
  I. Ristorcelli \inst{1}
  \and
  N. Ysard \inst{1,3}
} 
\institute{IRAP, Université de Toulouse, CNRS, UPS, 9 Av. du Colonel Roche, BP 44346, F-31028, Toulouse, cedex
  4, France
  \and
Space Telescope Science Institute, 3700 San Martin Drive, Baltimore,
MD 21218, USA
\and
Université Paris-Saclay, CNRS, Institut d'Astrophysique Spatiale, 91405, Orsay, France}

\authorrunning{Paradis et al.}
\titlerunning{}
\date{}
\abstract
{The {\it Planck} mission provided all-sky dust emission maps in the
  submillimeter (submm) to millimeter (mm) range at an angular resolution of 5$^{\prime}$. In addition, some specific sources can be observed at long
  wavelengths and higher
  resolution using ground-based telescopes. These observations are limited to small scales and are sometimes not
delivered to the community. These ground-based observations require
extensive data processing before they become available for scientific analysis, and
suffer from extended emission filtering. }
{At present, we are still unable to fully understand the
  emissivity variations observed in different astrophysical environments at long (submm and mm) wavelengths. Several models have been
  developed to reproduce the diffuse Galactic medium, and each
  distinct environment requires an adjustment of the models. It is
  therefore challenging to estimate any dust emission
  in the submm-mm at a better
  resolution than the 5$^{\prime}$ from {\it Planck}. 
  In this analysis, based on supervised deep
  learning algorithms, we produced dust emission predictions in
  the two {\it Planck} bands centered at 850 $\mic$ (353 GHz) and 1.38 mm (217 GHz) at the
  {\it Herschel} resolution (37$^{\prime\prime}$). Prediction or forecasting is a frequently used term  in machine learning or neural network research that refers to the output of an algorithm that has been trained on a given dataset and that is being used for modeling purposes.}
{{\it Herschel} data of Galactic
  environments, ranging from 160 $\mic$ to 500 $\mic$ and smoothed to an angular resolution of 5$^{\prime}$, were used to train the neural network. This
  training aimed to
  provide the most accurate model for reproducing {\it Planck} maps of dust
  emission at 850 $\mic$ and 1.38 mm. Then, using {\it Herschel}
  data only, the model was applied to predict dust emission maps at 37$^{\prime\prime}$.} 
{The neural network is capable of reproducing dust emission maps of various Galactic
  environments with a difference of only a few percent at the {\it Planck}
  resolution. Remarkably, it also performs well for nearby
  extragalactic environments. This could indicate that large
  dust grains, probed by submm or mm observations,
  have similar properties
  in both our Galaxy and nearby galaxies, or at least that their
  spectral behaviors are comparable in Galactic and extragalactic environments. For the first
  time, we provide to the community dust emission prediction maps at 850 $\mic$ and 1.38
  mm at the 37$^{\prime\prime}$ of several surveys: Hi-GAL, Gould
  Belt, Cold Cores, HERITAGE, Helga, HerM33es, KINGFISH, and Very
  Nearby Galaxies. The
  ratio of these two wavelength brightness bands reveals a 
  derived emissivity
  spectral index statistically close to 1 for all the surveys, which favors the hypothesis of a flattened
  dust emission spectrum for wavelengths larger than 850 $\mic$. }
{Neural networks appear to be powerful algorithms that are highly
    efficient at learning from large datasets and achieving accurate
    reproductions with a deviation of only a few percent. However, to
    fully recover the input data during the training, it is essential to sample a
    sufficiently large range of datasets and physical conditions. }


\maketitle
\section{Introduction}
Reproducing the global spectral behavior of dust emission throughout our
Galaxy from the infrared (IR) to the millimeter (mm) with dust models
has proven to be more difficult than envisioned. In this wavelength regime, the sky brightness ($I_{\nu}(\lambda)$) is often simply described by
a single modified black-body model, assuming a single dust temperature ($T$) and spectral index ($\beta$) along the line
of sight and an optically thin medium.  Even in the diffuse
interstellar medium (ISM) of our Galaxy, where the largest grains may
be thought to be at a single temperature and to have homogeneous properties, there is
no consensus yet on either the dust temperature determination or the exact spectral shape of
the dust emission in the far-infrared (FIR) to mm. The sub-millimeter (submm) behavior varies over the sky and its variations are not yet
understood. This behavior could be the result of a nonuniformity in the
emissivity power law through the
environments — in other words, $\beta$ variations — or the result of a break in the emissivity power law described by a
change in $\beta$ from the FIR to the submm (usually described as a
two modified black-body model). The
spectral behavior of the emissivity is not constant over the sky and
the origin of these variations is still uncertain.   
Several dust emission models have been developed in order to
reproduce the emission of the
large-scale diffuse ISM in our Galaxy \citep{Desert90, Draine07,
  Compiegne11, Jones13, Siebenmorgen23, Hensley23,
  Ysard24}. Unfortunately, the interpretation  of the dust emission spectrum relies on the dust model
being used. 
\begin{figure}
\begin{center}
\includegraphics[width=7.cm]{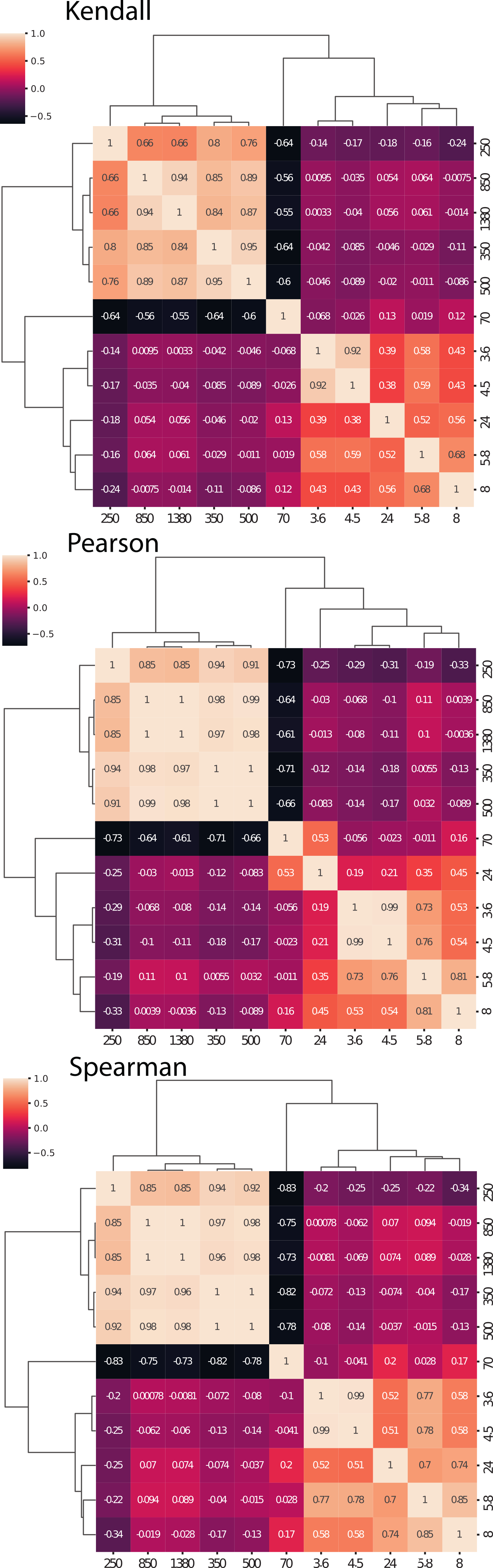}
\caption{Hierarchically clustered heat map of the different
  correlations between datasets (GLIMPSE at 3.6, 4.5, 5.8, and 8
  $\mic$; MIPSGAL at 24 $\mic$; Hi-GAL from 70 $\mic$ to 500 $\mic$, and
  {\it Planck} at 850 $\mic$ and 1.38 mm). All datasets have been
  smoothed to an angular resolution of 5$^{\prime}$ and normalized to the 160 $\mic$ data. Kendall, Pearson, and
  Spearman's correlation coefficients are given in each case.
  \label{corr}}
\end{center}
\end{figure}
At present, large-scale observations at long wavelengths ($>500$ $\mic$)
only exist at an angular resolution of 5$^{\prime}$, thanks to the {\it Planck}\footnote{{\it Planck} (http://www.esa.int/Planck) is a project of the
European Space Agency – ESA – with instruments provided by two
scientific consortia funded by ESA member states (in particular the lead
countries: France and Italy) with contributions from NASA (USA), and
telescope reflectors provided in a collaboration between ESA and a scientific Consortium led and funded by Denmark.}
mission \citep{Planck11}. It is therefore not
possible to investigate from large surveys the dust emission at small scales in this wavelength
range. Ground-based observatories from facilities such as ALMA and NOEMA, or
  instruments such as SCUBA-2, Bolocam, and NIKA2,
can provide continuum observations of dust at high angular resolution for specific sources at long wavelengths \citep[see for instance][]{Sadavoy13,
  Enoch06, Turner19, Katsioli23}. However, some of
the observations are not available to the community. In addition, the
reduction of the data is a complex task and most of the extended emission is
lost when removing the contribution from Earth's
atmosphere. Furthermore, to properly compare these observations with those of space-based
telescopes, a specific filtering should be applied to the on-board
satellite observations. Only a few experts are able to properly
perform this kind of processing \citep[][for example]{Roussel20}.

The submm and mm wavelength data are crucial to explore the behavior of
the dust emission spectrum. Using {\it Herschel}\footnote{{\it Herschel} is an ESA space observatory with science instruments
provided by European-led Principal Investigator consortia and with
important participation from NASA.}  data, \citet{Juvela15}
evidenced statistically an increase in the dust spectral index toward the coldest
regions of Galactic cold cores. They obtained values significantly
larger than the mean value of $\sim$1.6 obtained for the diffuse
Galactic emission \citep{PlanckXI}. \citet{Paradis14} also showed distinct trends in the dust emission
between cold and warm Galactic environments, with changes in the dust
emissivity index, using {\it Herschel} data. Several studies have also revealed
variations in the spectral index over large areas \citep[see for
instance][]{Paradis12b, PlanckXXIX}. 
Variations in $\beta$ across the
high-latitude sky between $\beta=1$ and $\beta=2$ with a 30$^{\prime}$
angular resolution have been reported, along with an optical
depth different than the one expected from uniform dust emission
properties. Grain coagulation from the diffuse to the dense medium could explain
the relative change in the dust emission behavior between
these two environments. Indeed, grain coagulation is
expected in cold regions due to the presence of ice mantle at the
surface of the grains \citep{Stepnik03, Kohler11, Kohler12, Jones13, Ysard15} and could
engender a change in the emissivity spectral shape. The two-level-system model \citep{Meny07,Paradis11}, which takes the
  physical aspect of amorphous dust material into account, could
  also explain the emissivity behavior. Indeed, dust in the ISM is 
  mostly amorphous, as is evidenced by the 10 $\mic$ silicate absorption
  profile in the region around Sgr A$^{\star}$ \citep{Kemper04}. The
  dust injected in the ISM after its formation in AGB stars is
  amorphous and crystalline in proportions that are not known. The
  absence of crystalline dust in the ISM probably results from the
  processing of the dust grain by cosmic rays and supernovae-generated
  shock waves.\\
  \citet{Draine12} argues that the
  flattening of the dust emission spectrum could be partly explained
  by a mixture of normal dust with a population of small magnetic nano-particles, such as metallic iron,
  magnetite Fe$_3$O$_4$, or maghemite $\gamma$-Fe$_2$O$_3$. However, the
  addition of spinning dust is required to account for the
  observed spectral energy distribution (SED) of the Small Magellanic Cloud (SMC) in the mm domain. Nevertheless, the results of
  \citet{PlanckXI} using
  polarization data from {\it Planck} do not favor this hypothesis. 
\begin{figure*}
\sidecaption
\includegraphics[width=12cm]{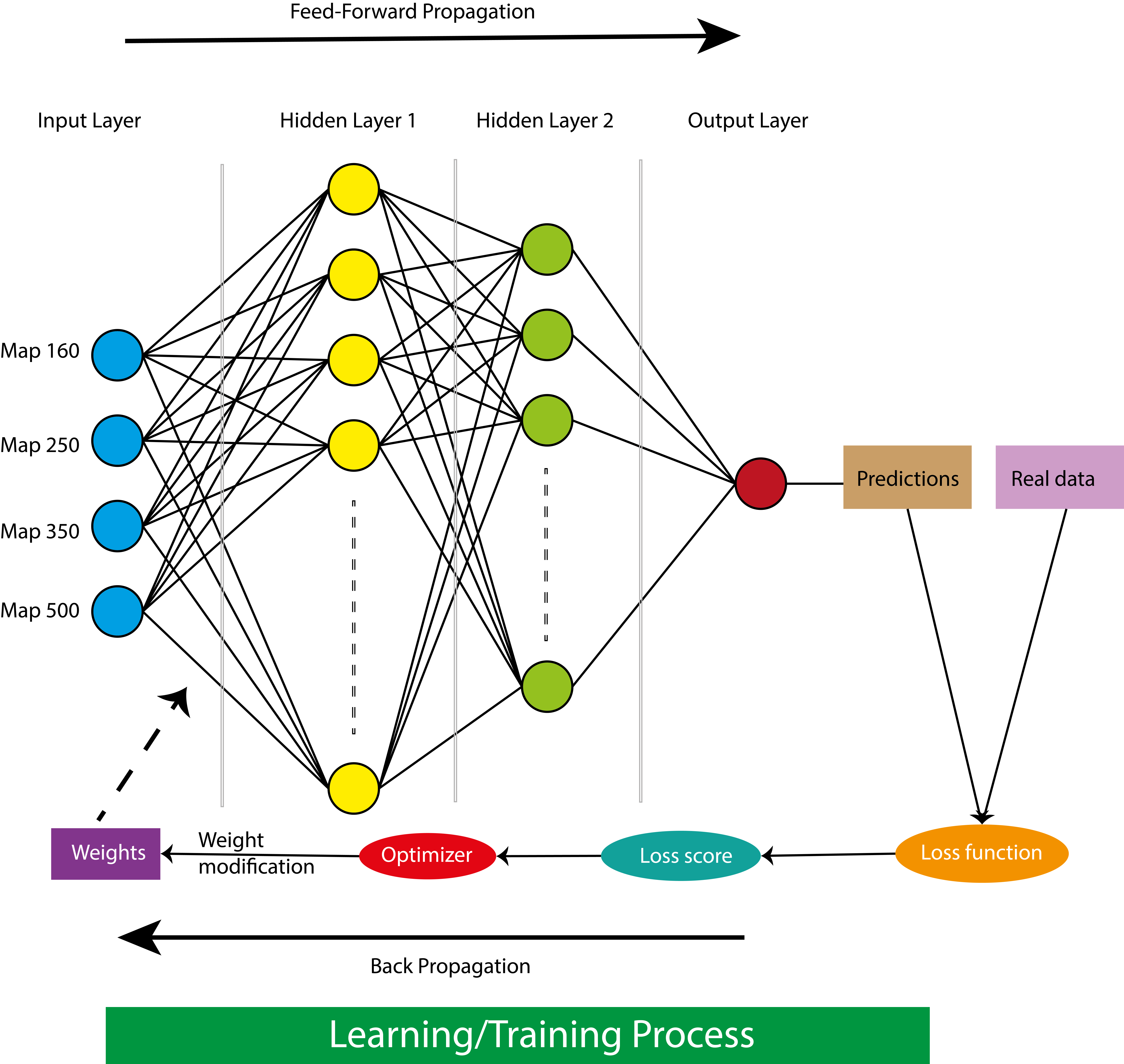}
\caption{Overview of a neural network for machine learning.}
\label{fig_neural_network} 
\end{figure*}

The resolution and sensitivity of {\it Herschel} data allowed the analysis of 
the ISM of more distant, but still close galaxies. \citet{Dale12}
analyzed 61 nearby galaxies as part of the KINGFISH program, and found that the low-metallicity galaxies 
in the sample are not colder than average but reveal an excess of
emission at 500 $\mic$ with
respect to the \citet{Draine07} model fits. Even if this behavior is
often detected in low-metallicity galaxies \citep{Izotov14} and the most significant excesses are 
detected in this type of environment \citep{Galametz11, Kirkpatrick13,Remy-Ruyer13}, the link between the
excess and metallicity has not been demonstrated yet. Long-wavelength
observations are crucial to explore the dust emission behavior. The {\it Planck}
satellite has been precious to get dust emission information at long
wavelengths, all over the sky. However, the combination of {\it Herschel} and {\it Planck}
data induces a loss of resolution in the {\it Herschel} data. In that case,
the SEDs of dust emission in external Galaxies are
reduced to the global brightness in the galaxies and do not allow one to
probe the ISM therein \citep[see for
instance][]{Hermelo16, Davies17, Tibbs18}.

The outline of this paper is to produce dust prediction maps in
the two {\it Planck} bands centered at 850
$\mic$ (353 GHz) and 1.38 mm (217 GHz) at the {\it Herschel} resolution (37$^{\prime\prime}$). Such maps 
could help us to understand the dust
emission spectrum variations, to carefully derive dust masses in various
environments, to help select targets to be observed at higher angular
resolutions, or to undertake foreground subtraction. The content of the paper is as follows. After a brief description of the datasets, we explore
the different correlations between multiwavelength data (Sect. \ref{sec_corr}). Then, we
describe the neural network methodology in Sect. \ref{sec_NN}, and
present different tools used in this analysis for manipulating HEALPix
data (Sect. \ref{sec_tools}). In Sect. \ref{sec_pred} we present
the prediction maps we
produced, in Sect. \ref{sec_disc} we discuss some interesting results, and in Sect. \ref{sec_cl} we
summarize this work.

\section{Data}
\label{sec_data}
\subsection{{\it Planck} and additional components}
\label{sec_planck_data}
We used the Planck/HFI 850 $\mic$ and 1.38 mm maps at an angular resolution of 5$^{\prime}$ \citep{Planck11}, corrected from the Zodiacal emission, available in the ESA
archive,\footnote{https://pla.esac.esa.int/} to trace the dust
emission. We used the {\it Planck} third release maps. Units in $K_{CMB}$ have
been converted to MJy/sr accordingly to \citet{PlanckIX}. We
subtracted the cosmic infrared background (CIB) monopole prediction by
removing 0.13 MJy/sr and 0.033 MJy/sr at 850 $\mic$ and 1.38 mm, as
is described in \citet{PlanckVIII}. We removed the CMB
contribution in each HFI map by using the {\it Planck} CMB map
reconstruction at 5$^{\prime}$ obtained from LGCMA,\footnote{see
  http://www.cosmostat.org/product/} a component separation
method. As is evidenced in \citet{PlanckXIII}, a total of nine lines could
contaminate the HFI channels. However, the authors concluded that only
the J=1-0, J=2-1 and J=3-2 CO lines could contribute to the total
emission, with significant transitions at 100 GHz (3mm) (50$\%$ of the
total emission in molecular clouds and in the Galactic plane) and 217 GHz
(1.38 mm) (15$\%$ of the total emission), weak at 353 GHz (850 $\mic$)
(less than 1$\%$), respectively. Hence, the CO
contamination is negligible in the other bands.The 217 GHz data and
353 GHz were
corrected for CO contamination using the 12CO (J=2-1) map
produced as part of the Commander multicomponent processing
\citep{Planck15Wehus}. Following \citet{Planck15Wehus}, free-free
  emission has a low impact at 217
GHz. \citet{Katsioli23} decomposed the emission of
the nearby edge-on galaxy NGC 891 at an angular resolution of 25$^{\prime \prime}$ into
dust, free-free, and synchrotron emission from 1 mm to 20 cm. They
found negligible free-free emission at 1.15 mm (less than 2$\%$). 
The resulting {\it Planck} maps therefore represent
the most reasonable description of the dust emission only. These are
the final maps we use in the following.
\begin{table}
  \caption{Central position of the JCMT/SCOPE sources used in this work.\label{table_regions}}
  \begin{center}
    \begin{tabular}{lcc}
\hline
      \hline
      Sources  & RA & Dec \\
      1 & 283.2 & 5.42 \\
      2 & 303.39 & 31.36 \\
      3 & 303.43 & 31.93 \\
      4 & 316.09 & 60.15 \\
      5 & 324.29 & 43.35 \\
      6 & 335.35 & 63.86 \\
      7 & 335.39 & 63.62 \\
       \hline
      \end{tabular}
  \end{center}
 \end{table} 

\subsection{{\it Herschel}}
\label{sec_herschel}
Thanks to the ``Centre d'Analyse de Données Etendues''
(CADE\footnote{https://cade.irap.omp.eu}), large maps of dust emission at different
wavelengths and resolution, in the HEALPix format,\footnote{https://healpix.jpl.nasa.gov/}
are available. For instance, the service delivers {\it Herschel} maps of the
different large programs at 37$^{\prime\prime}$ (NSIDE=16384),
1$^{\prime}$ (NSIDE=8192), and
4$^{\prime}$ (NSIDE=2048) angular resolutions, all of which we have produced
and made available on CADE. All {\it Herschel} data from this
analysis come from this service. We used HEALPix maps from 70
$\mic$ to 500 $\mic$ of the 
Hi-GAL \citep{Molinari10}, Galactic Cold Cores \citep{Juvela10}, and
Gould Belt \citep{Andre10} programs for training and testing the
neural network. We added maps of the HERITAGE \citep{Meixner10}, KINGFISH \citep{Kennicutt11}, Very Nearby
Galaxies \citep{Bendo12}, HerM33es \citep{Kramer10}, and Helga
\citep{Fritz12} programs to predict dust emission maps. 
We used data at a 37$^{\prime\prime}$ angular resolution to predict dust emission, and 4$^{\prime}$ angular
resolutions convolved to 5$^{\prime}$ to match the {\it Planck} resolution
in order to train and test the neural network. With full HEALPix files, which is
the case for the 4$^{\prime}$ files, the
smoothing was performed using the {\it smoothing} function of the
healpy Python package. 
\begin{figure}
\begin{center}
\includegraphics[width=9cm]{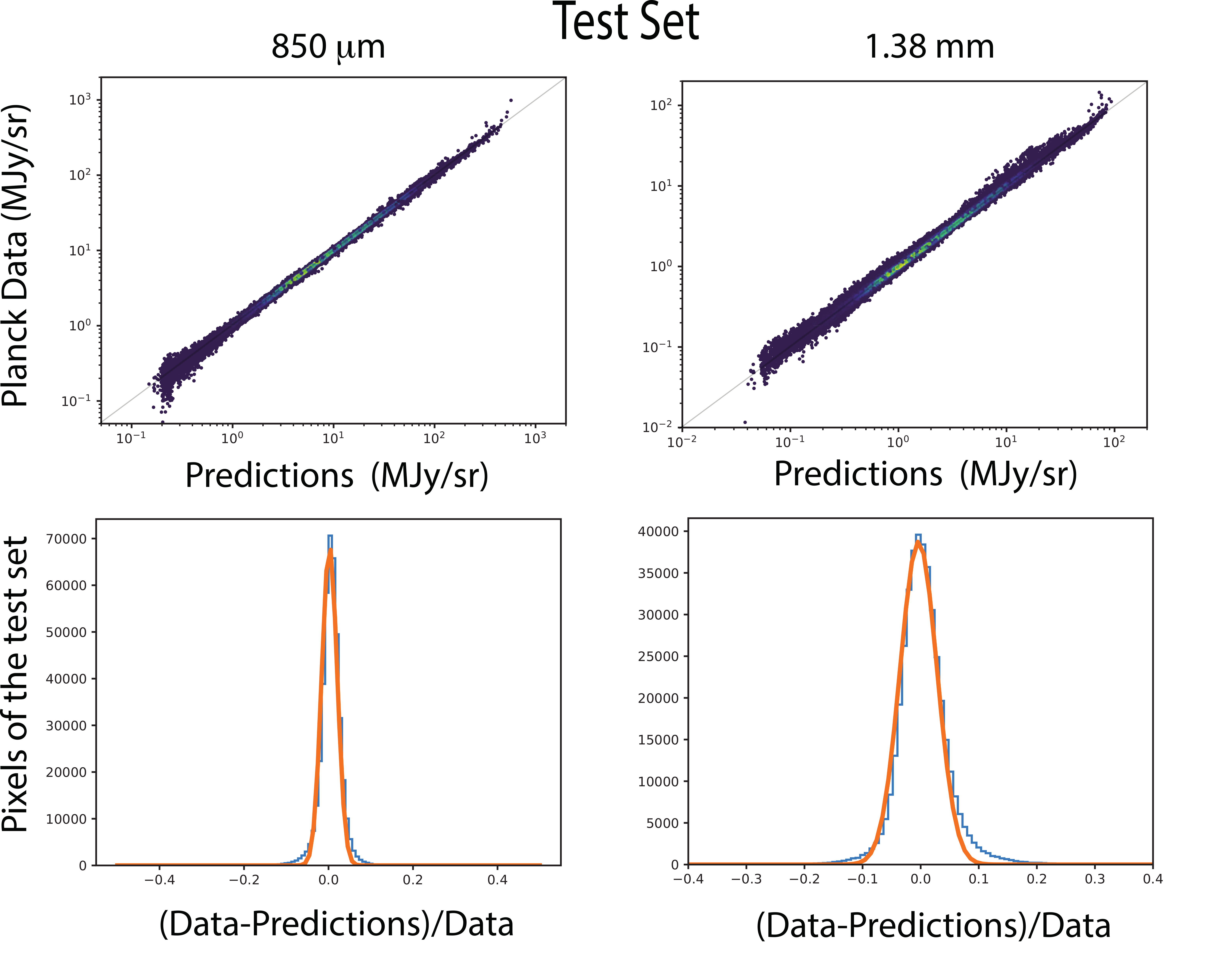}
\caption{Comparison between {\it Planck} and predictions performed on the
  test set (850 $\mic$ on the left and 1.38 mm on the right). Top: Correlation plots between {\it Planck} data and neural network
   predictions. Bottom: Histograms of the relative
  errors between the data and neural network predictions
  in blue, and Gaussian fits in orange. \label{corr_test} }
\end{center}
\end{figure}

\subsection{{\it Spitzer}}
Additional {\it Spitzer} data have been used before running the neural
network algorithms, especially for studying and evaluating any possible correlations between
the different datasets.  For this, we used HEALPix data from the two large
GLIMPSE programs (I/II/3D/360/Deep/Proper/APO surveys) combined with the
SMOG survey \citep{Carey08} at 3.6, 4.5, 5.8, and 8 $\mic$ \citep{Churchwell09}
and MIPSGAL at 24 $\mic$ \citep{Mizuno08}, taken from CADE, at an angular resolution of 4$^{\prime}$,
and then smoothed to 5$^{\prime}$. 

\subsection{SCUBA-2/JCMT}
To compare our 850 $\mic$ predictions at 37$^{\prime\prime}$ with observational
data, we used the {\it James Clerk Maxwell} Telescope (JCMT)/SCUBA-2 \citep{Holland13} 850 $\mic$
continuum observations as part of the Continuum Observations of
Pre-protostellar Evolution (SCOPE) program \citep{Eden19}. This
program consisted of observations of 1235 {\it Planck} Galactic cold clumps. We randomly
selected seven observations of this program from the EAO
archive\footnote{https://www.eaobservatory.org/jcmt/science/large-programs/scope/}
, which we converted from mJy/arcsec$^2$ to MJy/sr, and smoothed from
14.4$^{\prime\prime}$ to 37$^{\prime\prime}$. The coordinates of the
sources are provided in Table \ref{table_regions}. 
In addition, we considered SCUBA-2 850 $\mic$ observations of 
M31 obtained as part of the HASHTAG Program and delivered by the
consortium.\footnote{https://hashtag.astro.cf.ac.uk/DR1.html}.

\subsection{Bolocam}
\label{sec_bolocam}
We used Bolocam 1.1 mm data from the Bolocam Galactic Plane
Survey \citep[][BGPS]{Aguirre11}, at a resolution of
33$^{\prime\prime}$, in order to compare these with our 1.38 mm predictions. With a total coverage of 17 sq. deg, we used observations of
12 ultracompact HII regions (UCHIIs), already used
and described in \citet{Paradis14} (see their Table 1 for the central
position of the sources). Data in units of Jy/beam were
converted to MJy/sr using Eq. (16) from \citet{Aguirre11} before being
convolved to an angular resolution of 37$^{\prime\prime}$. However, it
is important to emphasize that Bolocam data suffer from noise, but
also from filtering, which induces an important loss of the extended
emission. For instance, 50$\%$ of the flux is attenuated for
structures extending to 3.8$^{\prime}$. For this reason, we did not
include data for cold cores, for which the significant extended
emission has been filtered, inducing a bad quality in the Bolocam data.
\begin{figure}
\begin{center}
  \includegraphics[width=9cm]{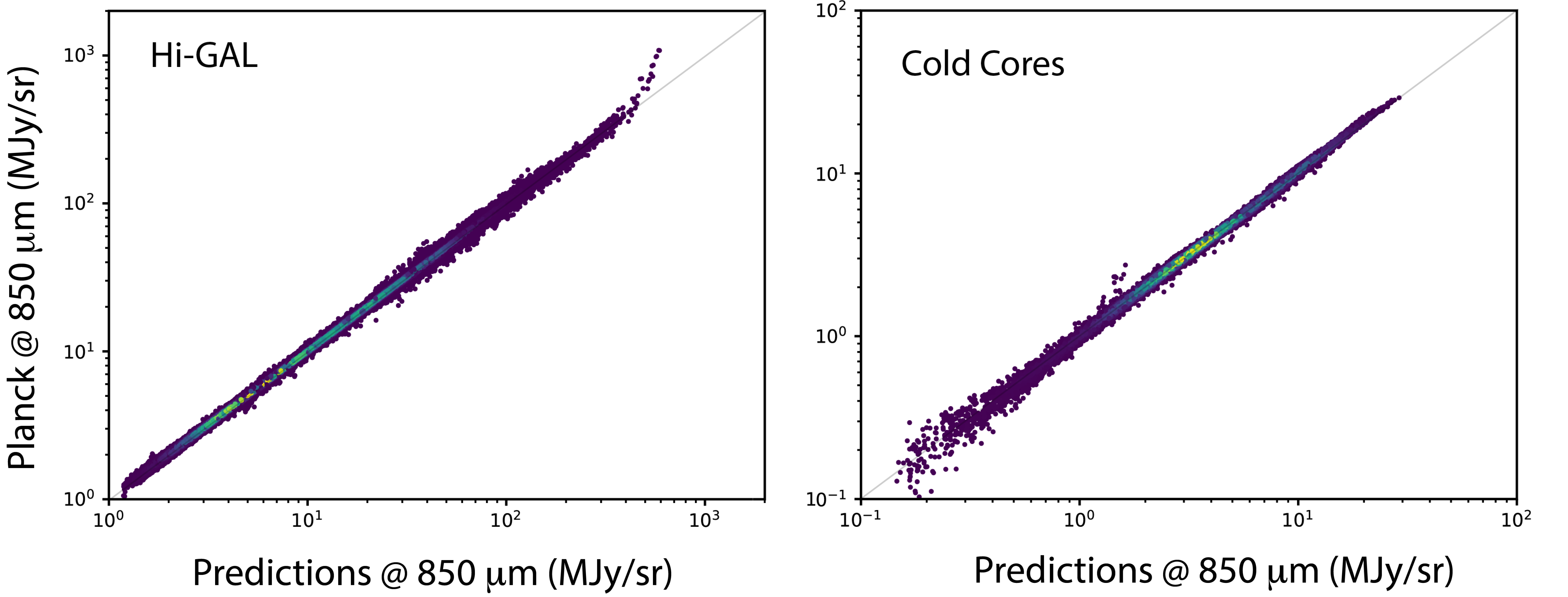}
\caption{Correlation plots between {\it Planck} data and neural network predictions at
  850 $\mic$, for the two Hi-GAL and Cold Cores {\it Herschel} large
  programs. See Fig. \ref{corr_850_app} for the correlation plots of
  all the {\it Herschel} large programs described in Sect. \ref{sec_herschel}.  \label{corr_850} }
\end{center}
\end{figure}
\begin{figure}
\begin{center}
\includegraphics[width=8.8cm]{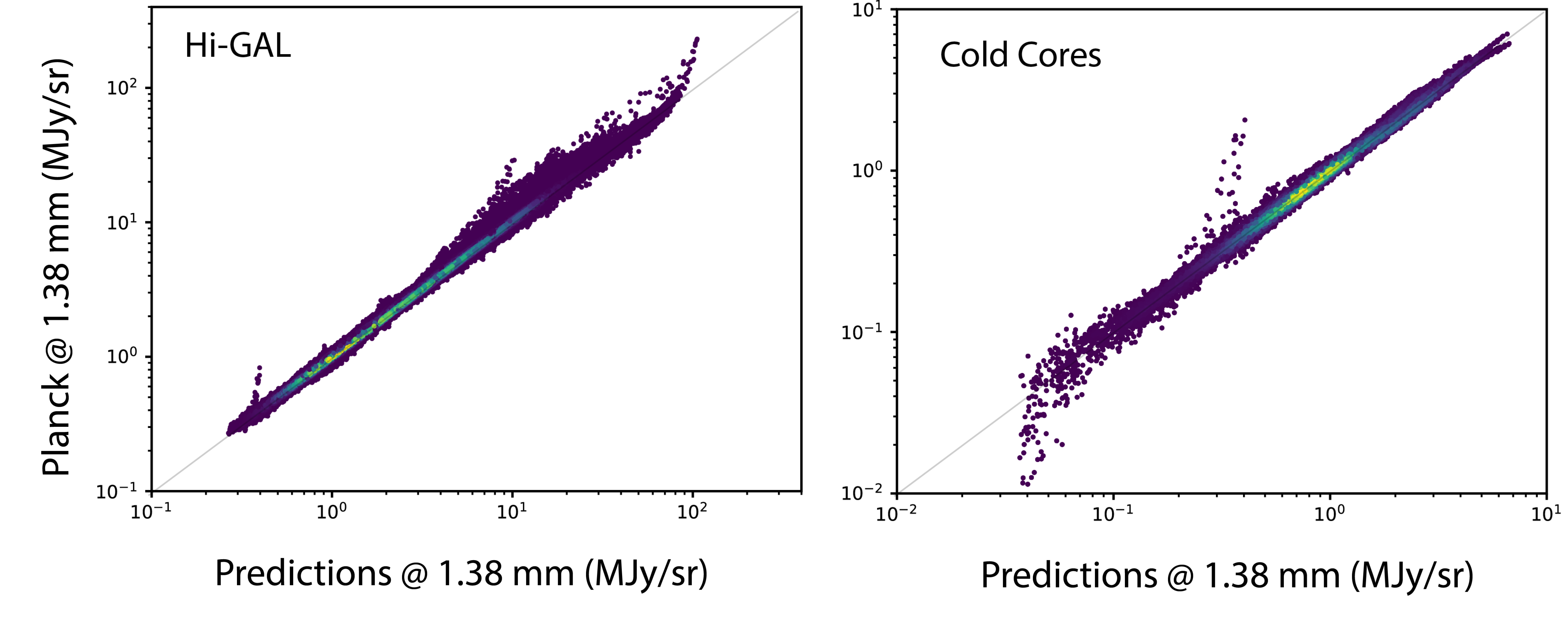}
\caption{Correlation plots between {\it Planck} data and neural network predictions at
  1.38 mm, for the two Hi-GAL and Cold Cores {\it Herschel} large
  programs. See Fig. \ref{corr_1p4_app} for correlation plots of all
  the {\it Herschel} large programs described in Sect. \ref{sec_herschel}. \label{corr_1p4} }
\end{center}
\end{figure}

\section{Correlations between multiwavelength infrared data}
\label{sec_corr}
To decide which wavelengths are required for the neural network to be able to predict dust
emission at 850 $\mic$ and 1.38 mm and which ones can be omitted, we
performed a correlation analysis.
Only the inner part of the Galactic
plane has been observed from the NIR to the FIR using {\it Spitzer} and
{\it Herschel} data. This analysis has therefore been performed
using the GLIMPSE (from 3.6 to 8 $\mic$), MIPSGAL (24 $\mic$), and
Hi-GAL (from 70 to 500 $\mic$) surveys, smoothed to an angular resolution of 5$^{\prime}$ (see Sect. \ref{sec_data}).
To evaluate the quality of the correlations, we
used the three following coefficients: Kendall (to measure the strength
of the dependence between two variables), Pearson (to measure the degree
of the relationship between linearly related variables), and Spearman
(to measure the degree of association between two
variables). The Pearson correlation is frequently used for normal
distributed data, whereas Spearman's and Kendall's coefficients are
suggested for non-normal data. The Kendall correlation is usually more robust and efficient
than the Spearman correlation, especially in the case of small samples or some outliers.
The results of the correlations are presented in Fig.
\ref{corr} for the three coefficients. We used the Python Seaborn
package to perform a hierarchically clustered heat map to facilitate the visualisation of
the correlations. Each case gives the correlation coefficient between
the data at two different wavelengths. Since the entire dust emission
spectrum increases with dust temperature, to remove this effect, which
could bias the correlations, all of the data were divided by the 160
$\mic$ brightnesses. For this reason, data at 160 $\mic$ do not appear in the
figure. Whichever coefficient correlation we inspect, the best
correlations (with values larger than 0.66 or 0.85 depending on the coefficient) are visible for wavelengths between 250 $\mic$ and
850$\mic$ or 1.38 mm. The coefficient values tend to decrease for
wavelengths below 70 $\mic$ (with values close to 0). Emission in the NIR and the underlying
processes do not affect or have a very low impact on the long-wavelength
emission. The 70 $\mic$ data are primarily dominated by
  emission from the large grain component, but they are often
  contaminated by emission from smaller grains, which weakens the
  correlation with longer wavelengths ($\lambda>$250 $\mic$). The negative correlation of the
  I$_{\nu}$(70)/I$_{\nu}$(160) ratios with
  the long-wavelength data arises because the 70 $\mic$ and 160 $\mic$ data lie on opposite sides of the peak of large grain emission. 
We
therefore only consider data from 160 $\mic$ to 500 $\mic$ in the
following to predict dust emission in the submm/mm using the neural network.

We note the absence of a correlation between the FIR and the MIR 24
$\mic$. Unfortunately, in the inner
Galactic plane covered by our observations, we do not have any cold clumps for which a decrease in the
small grain abundance is observed in our Galaxy, probably because they
stick to the largest grains or are not sufficiently heated in these
dense clumps and then do not radiate. However, this behavior has not been evidenced in the Large Magellanic Cloud (LMC) clumps, for
instance \citet{Paradis19}. Conversely, large grains should be
intensively heated to have a visible effect on the 24 $\mic$
emission. Such required high
temperatures ($>$80 K) of grains are probably not observed at an angular resolution of 5$^{\prime}$ because of dilution. 
\begin{figure*}
\begin{center}
\includegraphics[width=18cm]{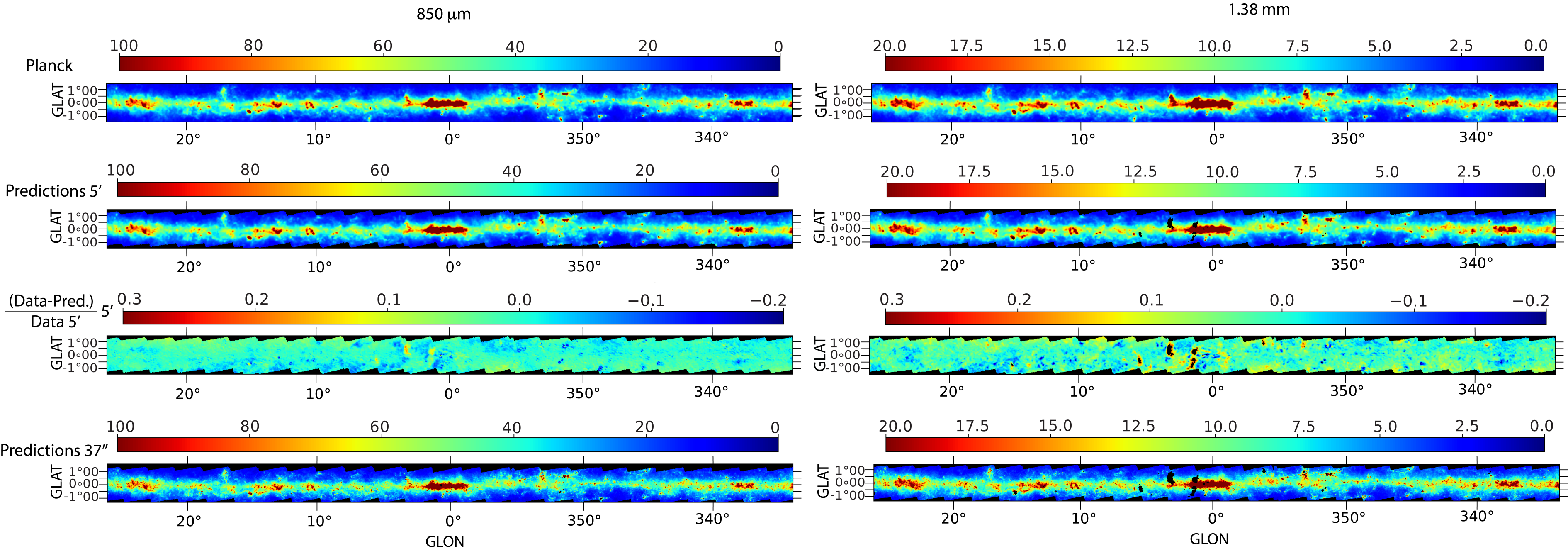}
\caption{Comparison between {\it Planck} data
  and predictions for a portion of the Galactic plane, at 850
  $\mic$ on the left and 1.38 mm on the right. From top to bottom: {\it Planck} data (5$^{\prime}$ angular
  resolution), neural network predictions
  (5$^{\prime}$), relative error between data and  predictions
  (5$^{\prime}$), and neural network predictions (37$^{\prime\prime}$)\label{fig_gal} }
\end{center}
\end{figure*}

\section{Neural network methodology}
\label{sec_NN}
To create our neural network, we used a multilayer perceptron model,
which consists of a supervised learning algorithm that learns a
function on a dataset using multiple layers. The multilayer
perceptron model is a feed-forward neural network with a structure
that consists of an input layer,
one or more hidden layers, and an output layer, an activation function,
and a set of weights and biases. The activation function determines
the output of a neuron based on its input, deciding whether a neuron
should be activated or not. A neural network without any activation function
becomes a linear regression model. Figure \ref{fig_neural_network} shows a simple
description of the neural network we use. The algorithm of our neural network consists
of two phases: \\
- forward propagation: the input is fed into the neural network and  the result is the output from the computations applied to
the data through the network.\\
- back-propagation: a supervised learning technique for training a neural network. The error between the predicted output and the actual output
is computed and propagated into the network. A gradient descent
optimization method is used to update the weights and
biases in order to reduce the error. \\
Our input layer consists of
four neurons that are the logarithms of the data at 160, 250, 350, and 500 $\mic$.  Our
two hidden layers process the information received from the input
layer. The output layer is the resulting logarithm of the prediction at 850 $\mic$ or 1.38
mm. 
The loss function of the neural network compares the target ({\it Planck}
data) and the predicted output values and measure the errors by giving a ``loss'' score. An
optimizer algorithm is used to update the weights and biases in order
to reduce the losses.
We split our data to get 2/3 of the pixels for training the neural network and 1/3
for testing. We used the $tanh$ activation function for all hidden
layers. We implemented a custom metric function defined as the
coefficient of determination R2, used the $Adam$ optimizer
with a learning rate of 0.001, and applied the $Huber$ loss
function. 

For training, the input corresponds to the {\it Herschel} Hi-GAL, Gould Belt, and Cold Cores
maps from 160 to 500 $\mic$, at a 5$^{\prime}$ angular
resolution. The output layer is the {\it Planck} 850 $\mic$ or
1.38 mm. We removed from the analysis any potential negative brightness
values in the maps. The strength of this work is the important number of pixels
resulting from the HEALPix format that combines a multitude of world coordinate system (WCS)
fit images and that allows one to perform such a statistical analysis.
Indeed, the neural network was applied on 1 242 681
independent pixels in total,
with 832596 pixels for training and 410085 pixels for testing. 
Input and output maps were normalized using the
$RobustScaler$ function.  We performed hyper-parameter tuning in order to obtain a
high-performing model, which results in only two hidden layers with 12
and 8 neurons in each, respectively, to predict the 850 $\mic$ output map, and 60
and 30 neurons with the 1.38 mm output map. Applying the dropout
regularization function to improve the performance of the model is not recommended in our neural network, since we do not
have a large number of nodes. 

\section{Tools for manipulating data in the HEALPix format}
\label{sec_tools}
All of the input data were in the HEALPix format, which is very useful to
analyze a large area on the sky. However, to visualize specific regions 
it can be easier to extract a WCS fits file. In addition, so far,
astronomers use their own codes to extract an SED, either using the HEALPix format or  the WCS one. 
The CADE service, created in 2012, provides astronomical data production in the
HEALPix format at different resolutions, data archiving, and dissemination to the community. All the
data are made to be Virtual Observatory-compatible, through the HiPS format
\citep{Fernique15}. In
addition to the database, the service offers different tools for
manipulating data in the HEALPix format:\\
-  The drizzling software library\footnote{http://cade.irap.omp.eu/dokuwiki/doku.php?id=software} (in Python), which reprojects data
from HEALPix to the local
WCS, \\
-  A web interface of the Drizzlib (DrizzWeb),\footnote{http://drizzweb.irap.omp.eu/}
\\
- The first SED extractor tool\footnote{http://cade.irap.omp.eu/dokuwiki/doku.php?id=sed$\_$extractor} (in Python), which extracts an SED from HEALPix, WCS, or a mixture of HEALPix-WCS fits files.\\
In the following, we give a brief description of each tool.

\subsection{World coordinate system fits file extraction}
The first version of Drizzlib has been developed in the Interactive
Data Language (IDL), in the framework of the {\it Planck} mission, to manipulate
HEALPix data. The advantage of this format is that it has a unique
pixelization over the entire sky, with each pixel covering the same
surface area. 
The
description of the method for the transformation from WCS fits files to HEALPix files is given in \citet{Paradis12a} (see Appendix
A). The
method works in exactly the same way in the inverse transformation.

The IDL Drizzlib version has been replaced by a Python version, usable
for all projections and coordinate systems. With the improvement of the
resolution, the Drizzlib has been extensively modified to be able to
work with HEALPix files with large NSIDE ($>$8192). A simplified
web interface of the Drizzlib is also available online. These are the
only tools able to manipulate large HEALPix files, ensuring the flux
conservation from one pixelization to the other. We use a mosaicking
method that computes the surface of the pixel intersections, and use these values as weights to
extract the WCS fits file from the HEALPix format. 

In this work, Drizzlib and DrizzWeb have extensively been used to
extract WCS fits files in specific regions of the HEALPix prediction
maps, for comparison with other maps or to
produce an SED. The extraction time depends on the NSIDE of the
original HEALPix map, as well as the size of the output WCS fits
map. In the case of NSIDE=16384 and for output regions of several
degrees (such as the Magellanic Clouds and part of the Hi-GAL region),
the process takes too long. In that case, we generated HiPS of the
prediction maps and extracted the WCS file using the Aladin tool. This
process is acceptable for visualization only and not for data analysis,
since the flux conservation is not guaranteed.   
\begin{figure}
\begin{center}
\includegraphics[width=9cm]{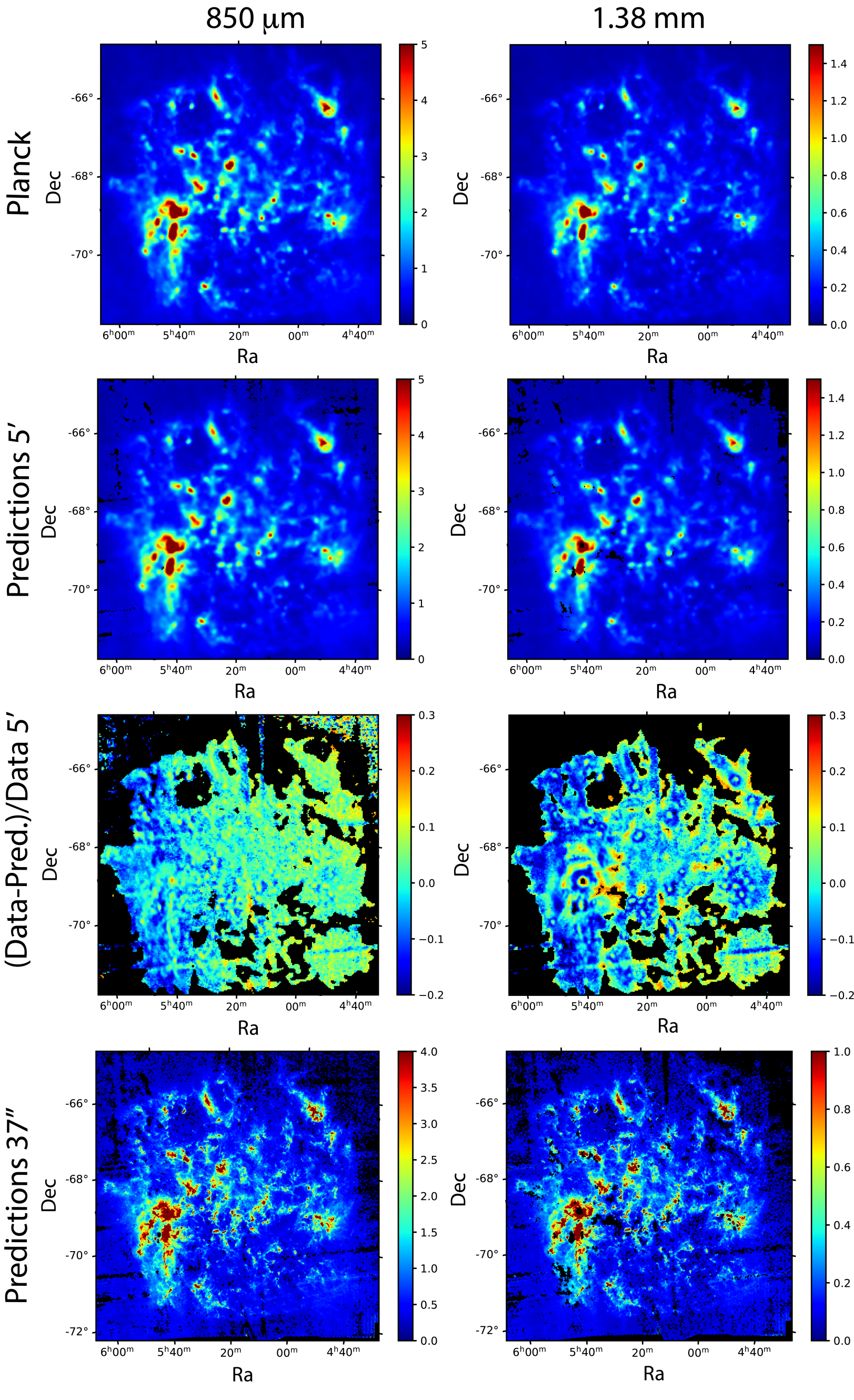}
\caption{Comparison between {\it Planck} data
  and predictions for the LMC, at 850
  $\mic$ on the left and 1.38 mm on the right. From top to bottom: {\it Planck} data (5$^{\prime}$ angular
  resolution), neural network predictions
  (5$^{\prime}$), relative error between data and predictions
  (5$^{\prime}$), and neural network predictions (37$^{\prime\prime}$).\label{fig_lmc} }
\end{center}
\end{figure}

\subsection{Spectral energy distribution extraction}
\label{sec_sed_extract}
All the SEDs presented in the following sections have been produced
using the SED extractor. The SED extractor offers large flexibility in the method of
extraction. The interest of such a tool is to accept WCS fits and/or
HEALPix files as inputs. The conversion from one format to the other is based on
the Python Drizzlib code, which ensures flux conservation. In addition,
we can specify different shapes of extraction regions, allowing cuts
in the coordinates limit, cuts in units of the maps, but
also mixing the different possibilities.
Further developments of this tool will improve the speed of the format
conversion and allow for correlation plots.

We used the same regions as in
\citet{Paradis14} for extractions using {\it Herschel}-Bolocam data and
the prediction maps; that is, using circles of 27.8$^{\prime\prime}$
centered on the source to compute the
average brightness, and a
circle annulus extended up from 27.8$^{\prime\prime}$ to
55.6$^{\prime\prime}$ to compute the median brightness in the background. As opposed to
\citet{Paradis14}, in which the SEDs were generated in Jy using
aperture photometry codes, we extracted SEDs in MJy/sr. The method of
extraction is different as well as the considered pixels in the region
and background, since some pixels are masked in the prediction maps
(if relative errors are larger than 20$\%$ in the 5$^{\prime}$
maps). To compute the SED using {\it Herschel}-JCMT data and prediction maps, we used the same circle radius, as well as a circle annulus for
the background.

In the case of M31,  we produced the SED of the galaxy by averaging the brightnesses inside an
ellipse centered at (l,b) =  (121.3; -20.9) and defined
by a width of 0.55$\degr$, height of 2$\degr$, and angle inclination of
38$\degr$. We removed a background
corresponding to the median brightness into a circle of 0.1$\degr$ in
radius centered at (l,b) = (121.3; -20.9).

We considered calibration uncertainties of 7$\%$ for {\it Herschel}
\citep[see][for PACS 160 $\mic$, and the observer manual v2.4 for SPIRE]{Balog14}, 10$\%$ for SCUBA-2 \citep{Jenness02},
20$\%$ for Bolocam \citep[][]{Ginsburg13}, and 4$\%$ and 7$\%$ for
 predictions at 850 $\mic$ and 1.38 mm (see Sect. \ref{sec_pred_5arcm}). 

\section{Dust emission predictions at 850 $\mic$ and 1.38 mm}
\label{sec_pred}
After training the neural network, we first used as inputs the {\it Herschel} maps that we
converted to the HEALPix format and made available on the CADE
service. The maps have been smoothed to 5$^{\prime}$ (see
Sect. \ref{sec_herschel}), to be compared with the {\it Planck}
dust emission data. Then, we used the {\it Herschel} HEALPix maps at an angular resolution of 37
$^{\prime \prime}$ to predict dust emission
in the {\it Planck} bands centered at 850
$\mic$ and 1.38 mm.

\subsection{Prediction maps at 5$^{\prime}$}
\label{sec_pred_5arcm}
We applied the best models derived from our neural networks to the Hi-GAL, Gould
Belt, and Cold Cores HEALPix maps at 5$^{\prime}$. We compare in Fig.
\ref{corr_test} the predicted output and the {\it Planck} maps, restricted
to the pixels used in the test set. The correlation plots show that
the predicted data reproduce well the real data. There is clearly a
low bias and a low variance at both wavelengths. The histograms of
relative errors (see
Fig. \ref{corr_test}) indicate slightly larger relative errors for the 1.38 mm than the 850
$\mic$. This point is not surprising, since emission at 1.38 mm
can still be slightly contaminated
by other emission features in the dust continuum, even after CMB, CIB,
and CO removal (see Sect. \ref{sec_planck_data}). These
  potential contaminations can induce more complexity to
recover the {\it Planck} dust emission data. However, for both wavelengths,
the histograms are centered on values close to 0 (0.0023 and -0.0037
at 850 $\mic$ and 1.38 mm) and have small standard
deviations ($\simeq$ 4$\%$ and 7$\%$ at 850 $\mic$ and 1.38 mm for
95$\%$ confidence interval).
\begin{figure}
\begin{center}
\includegraphics[width=9cm]{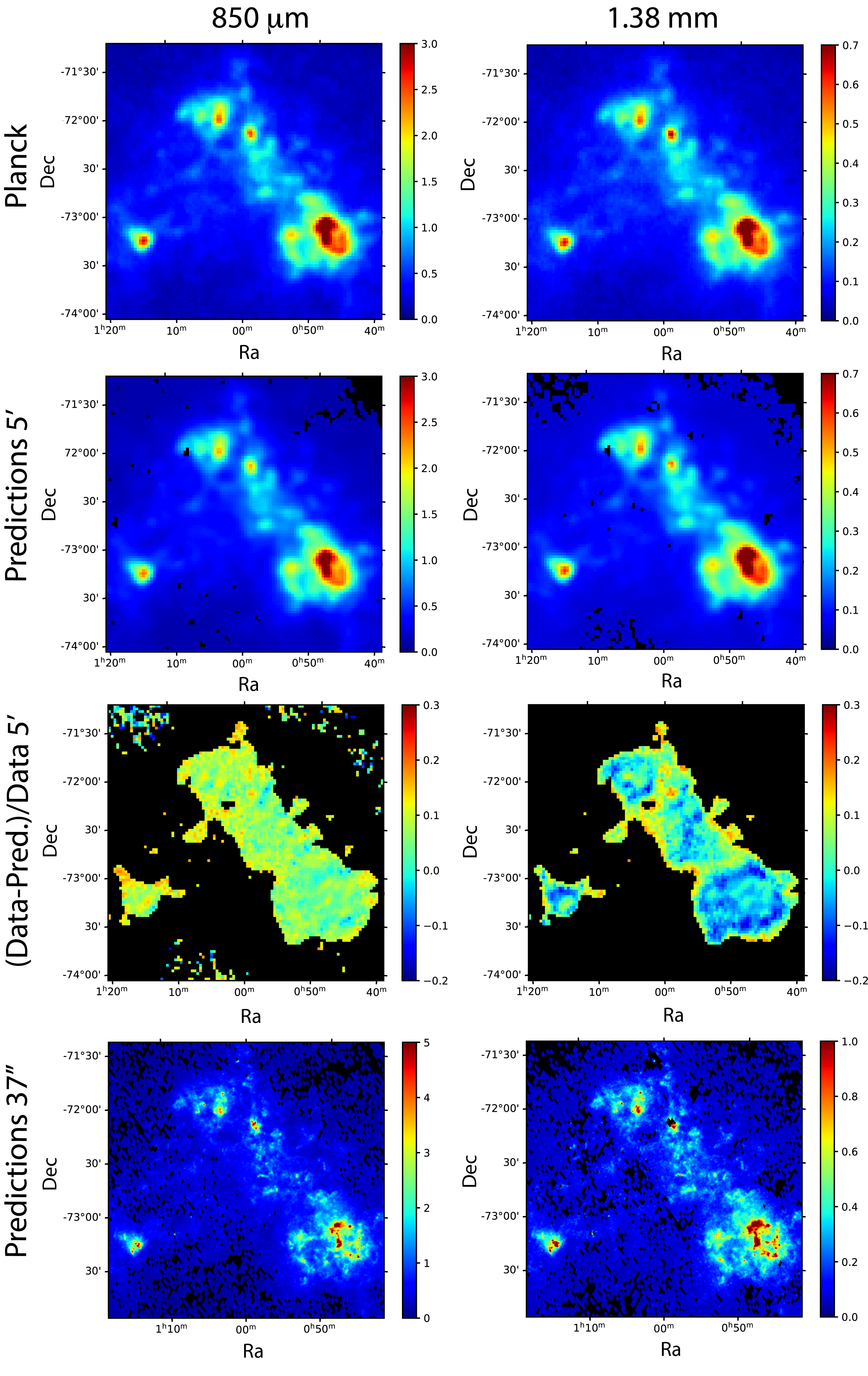}
\caption{Comparison between {\it Planck} data
  and predictions for the SMC, at 850
  $\mic$ on the left and 1.38 mm on the right. From top to bottom: {\it Planck} data (5$^{\prime}$ angular
  resolution), neural network predictions
  (5$^{\prime}$), relative error between data and predictions
  (5$^{\prime}$), and neural network predictions (37$^{\prime\prime}$). \label{fig_smc} }
\end{center}
\end{figure}

We show the correlations for the Hi-GAL and Cold Cores programs in
Figs. \ref{corr_850} and \ref{corr_1p4}, and for all the {\it Herschel}
programs described in Sect. \ref{sec_herschel}  in Figs. \ref{corr_850_app} and
\ref{corr_1p4_app}. We removed very low-brightness
pixels in the correlation plots, for all the external galaxy surveys
(HERITAGE, KINGFISH, Helga, HerM33es, and VNGS). Indeed, these pixels
are dominated by the noise in the data, for a  brightness level below
0.05 to 0.1 MJy/sr at 1.38 mm, depending on the surveys.
Again, correlations appear slightly
better at 850 $\mic$ than 1.38 mm. However, we observe a systematic
trend: predictions for high brightness can be
underestimated (mainly in the Hi-GAl, Gould Belt, and HERITAGE
programs). These very bright regions correspond to specific UCHII regions
(Galactic center and Orion regions for the Milky Way (MW), and 30-Doradus
for the LMC, for instance). However, these discrepancies between data
and predictions only concern a few pixels (less than 0.2$\%$ in
the Hi-GAL program at 1.38 mm, for instance), and are not even visible in
the relative error histograms. However, to be rigorous, in the
delivered maps, we removed
pixels with relative errors exceeding 20$\%$. These same regions
have been removed in the prediction maps at an angular resolution of 37 $^{\prime \prime}$. The other UCHII regions do not evidence any
bias in the prediction maps. This impossibility of the neural network reproducing the
very bright pixels could come from two reasons. The first one
  is that very bright pixels represent only a few pixels and in that
  case the statistics may not be sufficient for the neural network to be
  correctly trained on these pixels. Secondly, the dust emission in
  these very bright masked pixels may exhibit behavior distinct from
  other bright regions. One possibility is a low but non-negligible
  contamination by free-free emission in these areas, particularly at
  1.38 mm. We tried to train a distinct neural network for these specific
pixels but the low statistics did not allow us to get better results. We
therefore preferred to remove these pixels from the prediction maps. 

\begin{figure}
\begin{center}
\includegraphics[width=9cm]{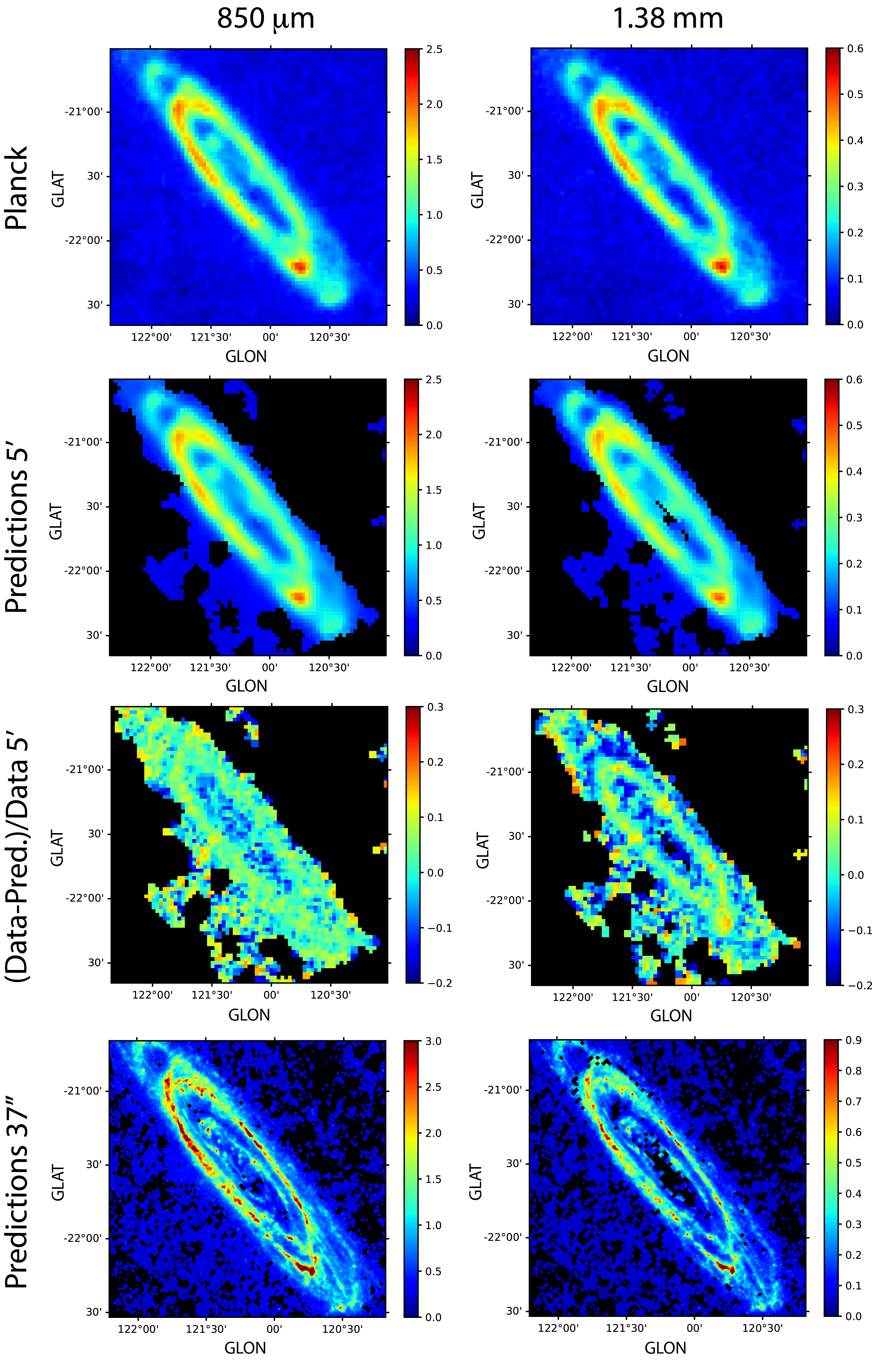}
\caption{Comparison between {\it Planck} data
  and predictions for M31, at 850
  $\mic$ on the left and 1.38 mm on the right. From top to bottom: {\it Planck} data (5$^{\prime}$ angular
  resolution), neural network predictions
  (5$^{\prime}$), relative error between data and predictions
  (5$^{\prime}$) and neural network predictions (37$^{\prime\prime}$). \label{fig_m31} }
\end{center}
\end{figure}

\begin{figure}
\begin{center}
\includegraphics[width=9cm]{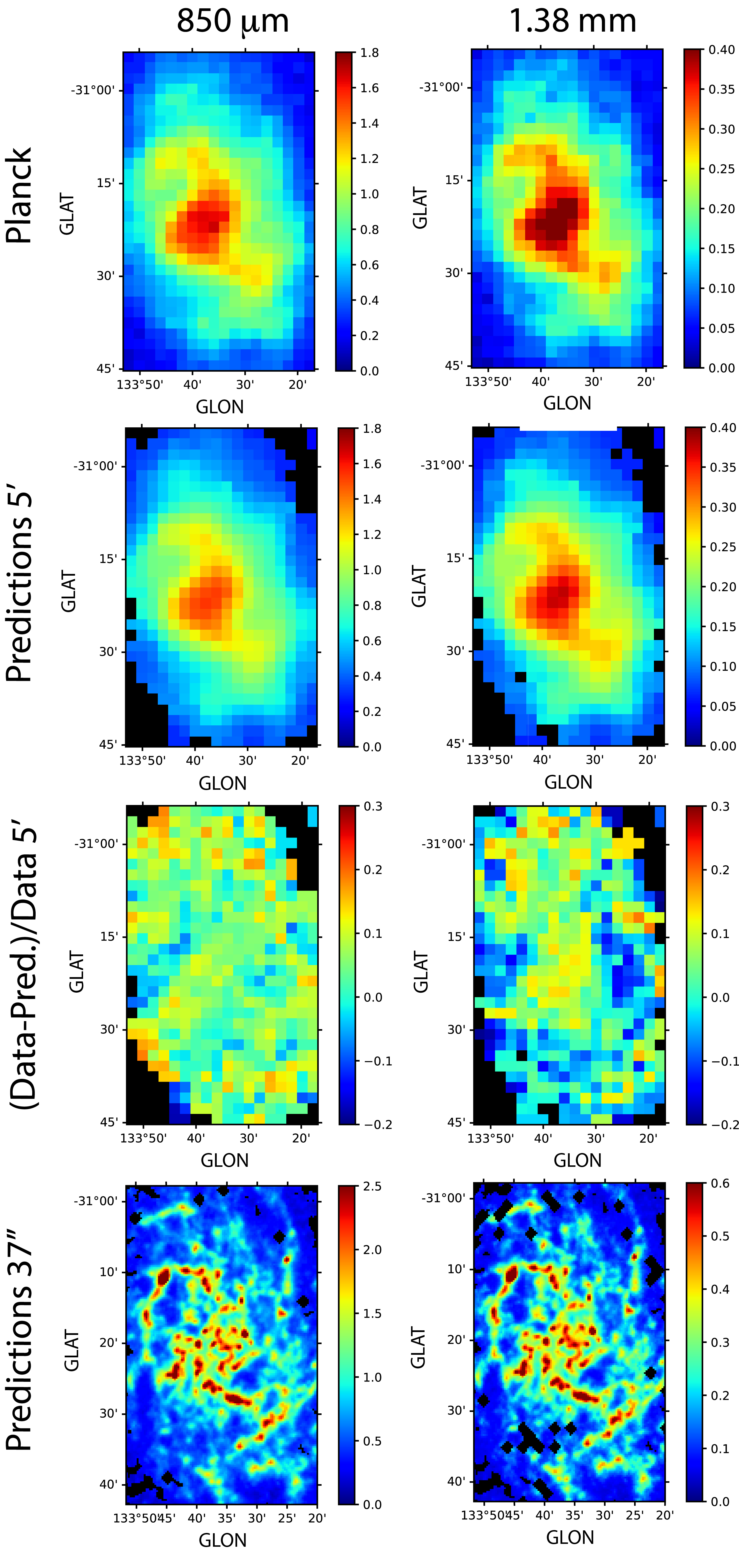}
\caption{Comparison between {\it Planck} data
  and predictions for M33, at 850
  $\mic$ on the left and 1.38 mm on the right. From top to bottom: {\it Planck} data (5$^{\prime}$ angular
  resolution), neural network predictions
  (5$^{\prime}$), relative error between data and predictions
  (5$^{\prime}$) and neural network predictions (37$^{\prime\prime}$). \label{fig_m33} }
\end{center}
\end{figure}

\begin{figure}
\begin{center}
\includegraphics[width=9cm]{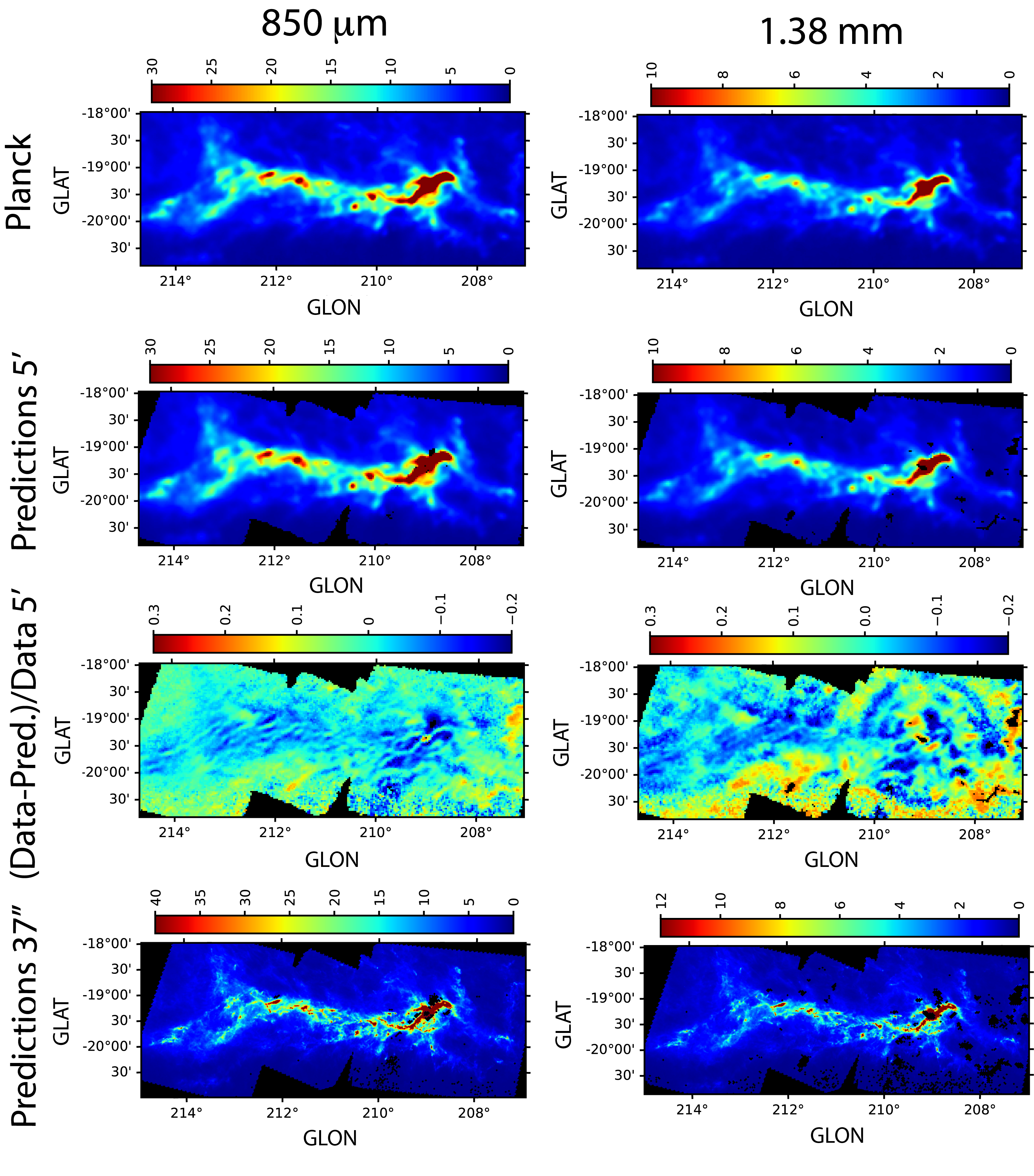}
\caption{Comparison between {\it Planck} data
  and predictions for the Orion region, at 850
  $\mic$ on the left and 1.38 mm on the right. From top to bottom: {\it Planck} data (5$^{\prime}$ angular
  resolution), neural network predictions
  (5$^{\prime}$), relative error between data and predictions
  (5$^{\prime}$) and neural network predictions (37$^{\prime\prime}$). \label{fig_orion} }
\end{center}
\end{figure}

\begin{figure}
\begin{center}
\includegraphics[width=9cm]{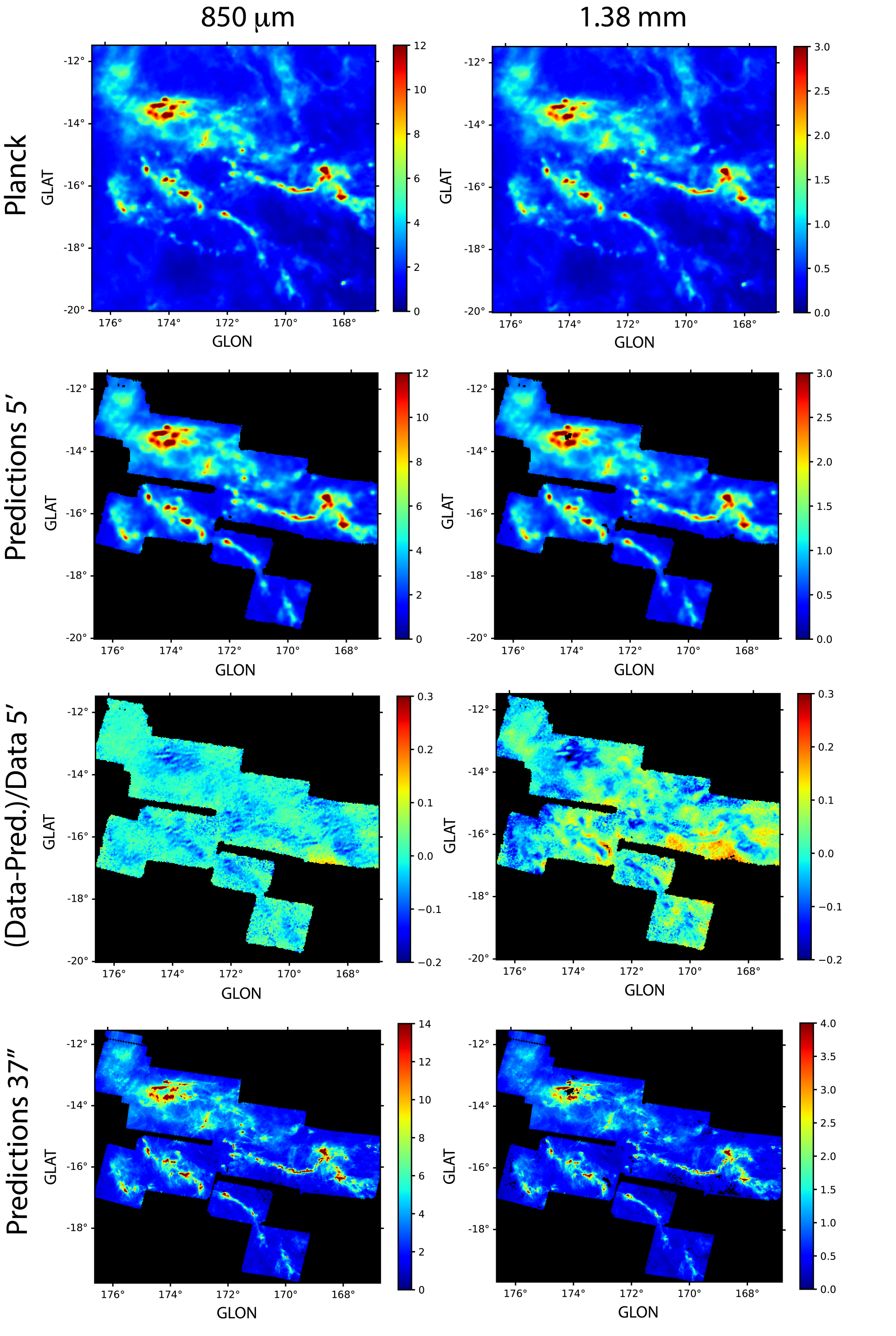}
\caption{Comparison between {\it Planck} data
  and predictions for the Taurus region, at 850
  $\mic$ on the left and 1.38 mm on the right. From top to bottom: {\it Planck} data (5$^{\prime}$ angular
  resolution), neural network predictions
  (5$^{\prime}$), relative error between data and predictions
  (5$^{\prime}$) and neural network predictions (37$^{\prime\prime}$). \label{fig_taurus} }
\end{center}
\end{figure}

\begin{figure}
\begin{center}
\includegraphics[width=9cm]{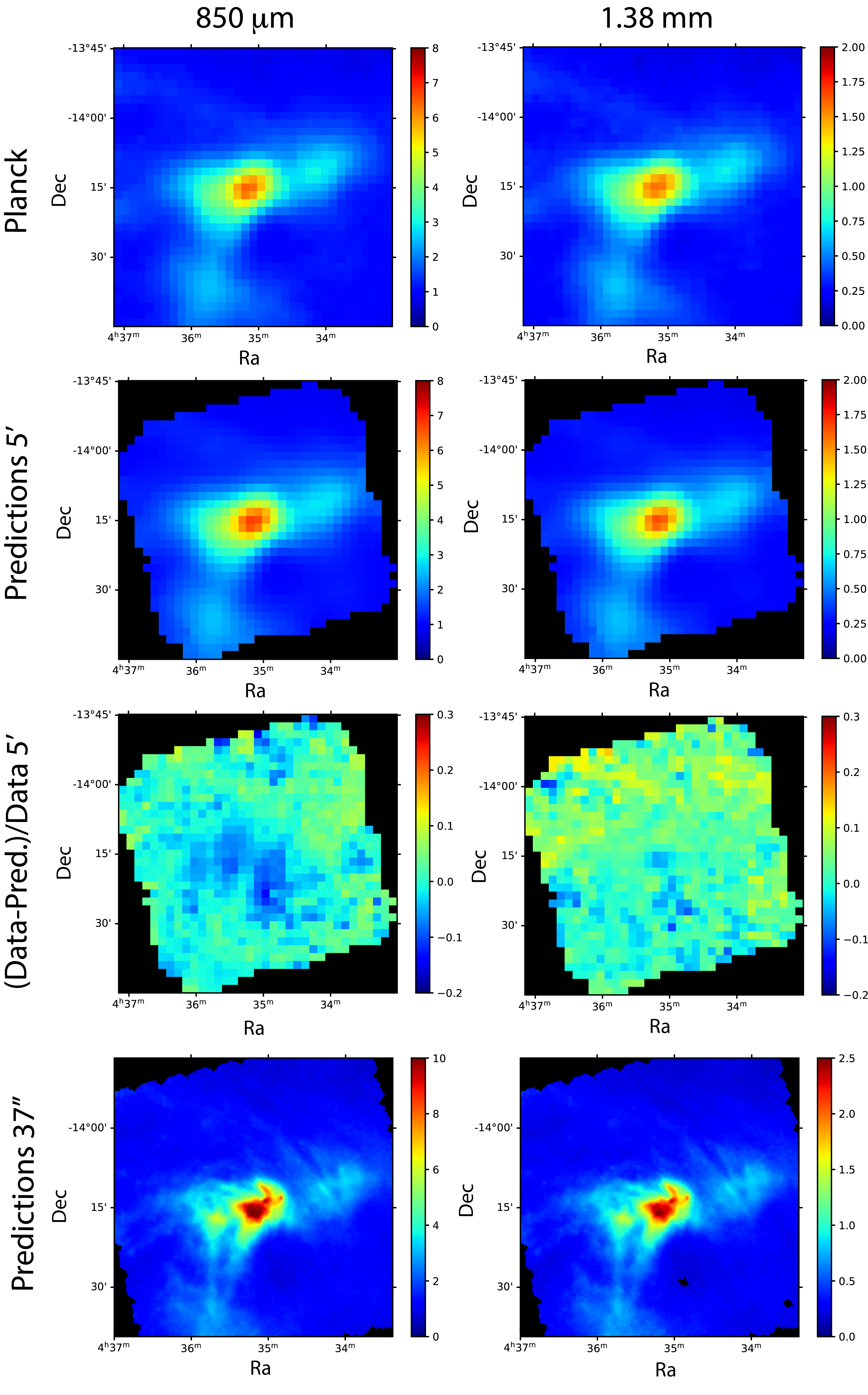}
\caption{Comparison between {\it Planck} data
  and predictions for LDN1642, at 850
  $\mic$ on the left and 1.38 mm on the right. From top to bottom: {\it Planck} data (5$^{\prime}$ angular
  resolution), neural network predictions
  (5$^{\prime}$), relative error between data and predictions
  (5$^{\prime}$) and neural network predictions (37$^{\prime\prime}$). \label{fig_ldn1642} }
\end{center}
\end{figure}

\begin{figure}
\begin{center}
\includegraphics[width=9cm]{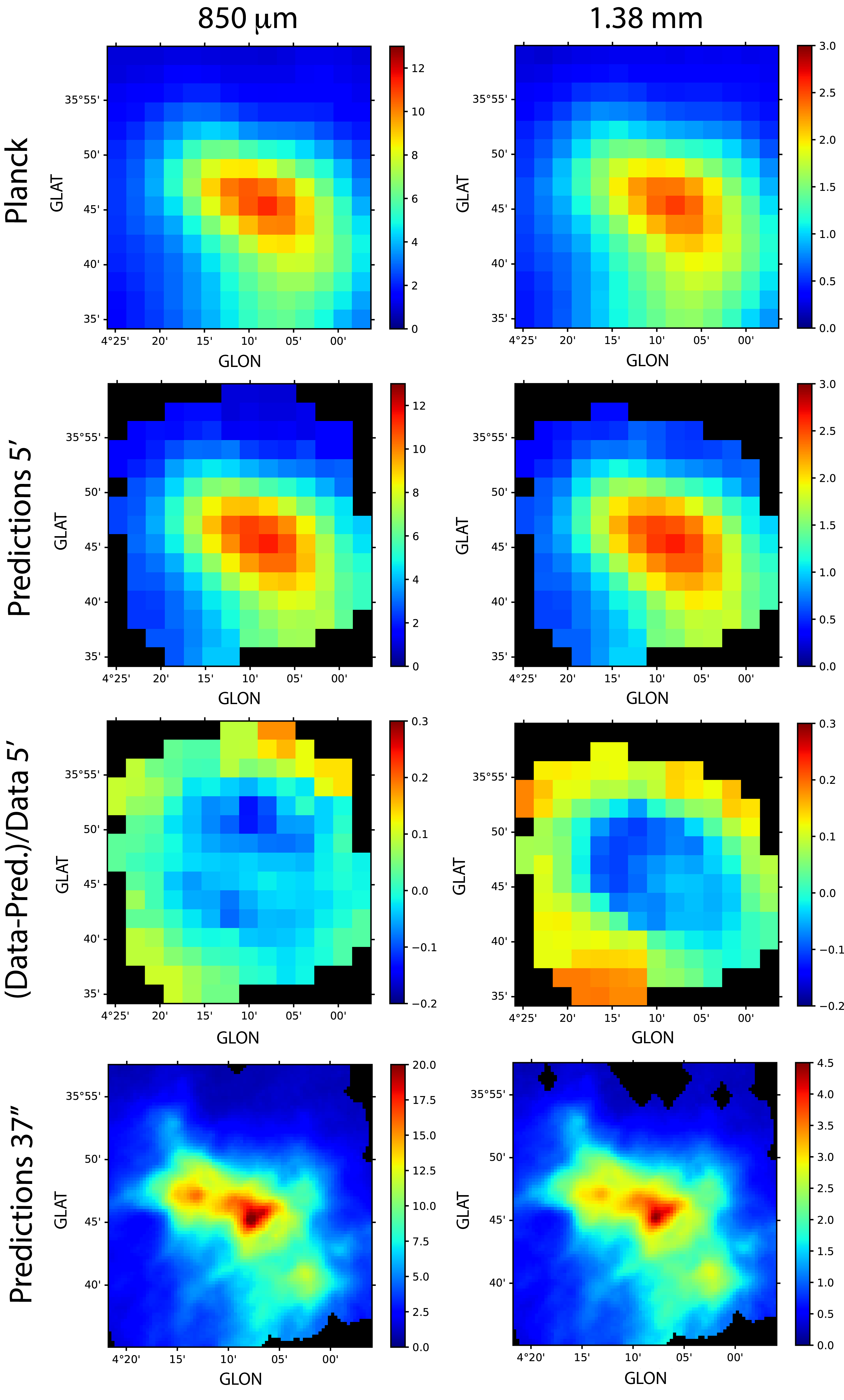}
\caption{Comparison between {\it Planck} data
  and predictions for L134, at 850
  $\mic$ on the left and 1.38 mm on the right. From top to bottom: {\it Planck} data (5$^{\prime}$ angular
  resolution), neural network predictions
  (5$^{\prime}$), relative error between data and predictions
  (5$^{\prime}$) and neural network predictions (37$^{\prime\prime}$).\label{fig_l134} }
\end{center}
\end{figure}

\begin{figure}
\begin{center}
\includegraphics[width=9cm]{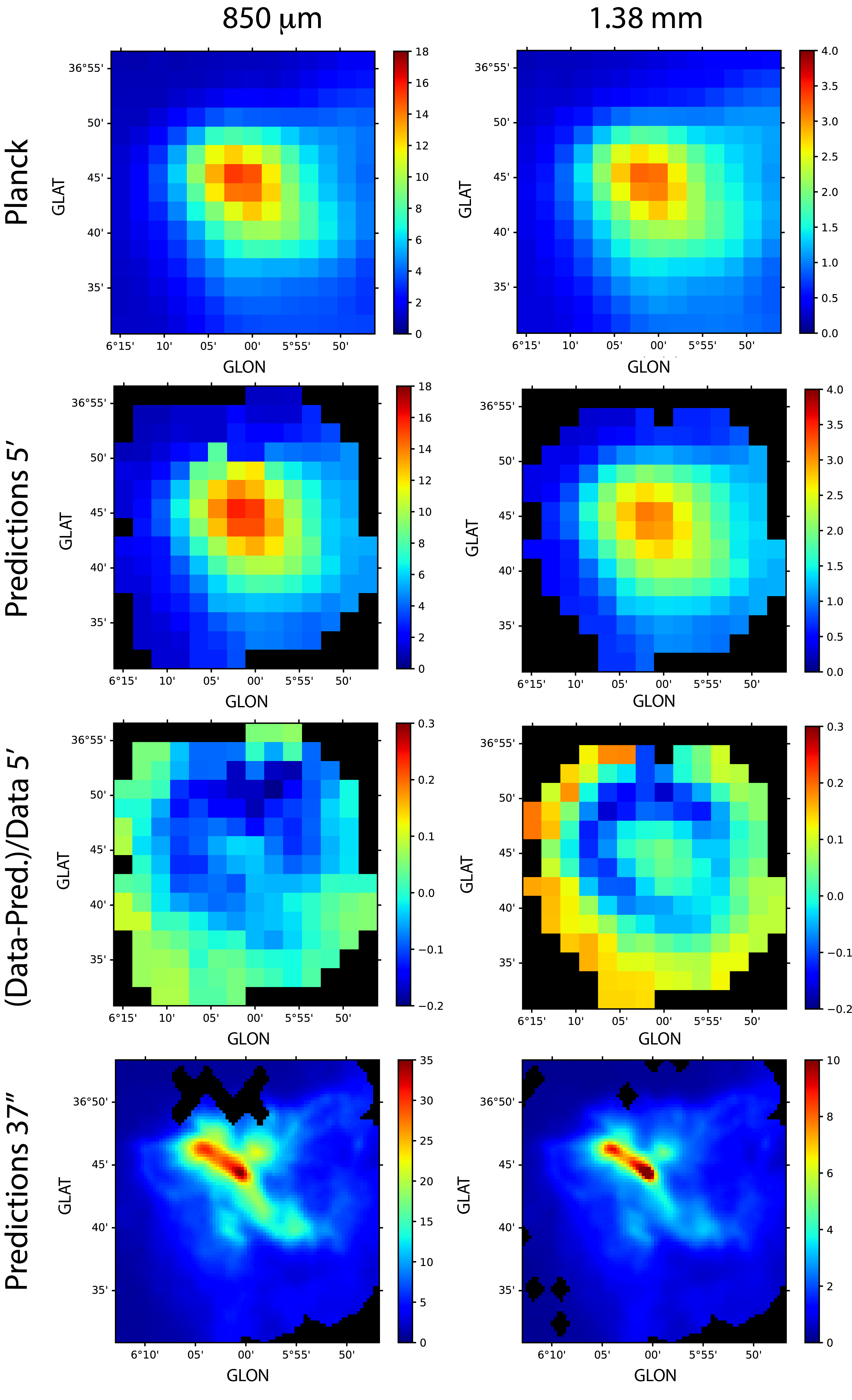}
\caption{Comparison between {\it Planck} data
  and predictions for L183, at 850
  $\mic$ on the left and 1.38 mm on the right. From top to bottom: {\it Planck} data (5$^{\prime}$ angular
  resolution), neural network predictions
  (5$^{\prime}$), relative error between data and predictions
  (5$^{\prime}$) and neural network predictions (37$^{\prime\prime}$). \label{fig_l183} }
\end{center}
\end{figure}

\begin{figure}
\begin{center}
\includegraphics[width=9cm]{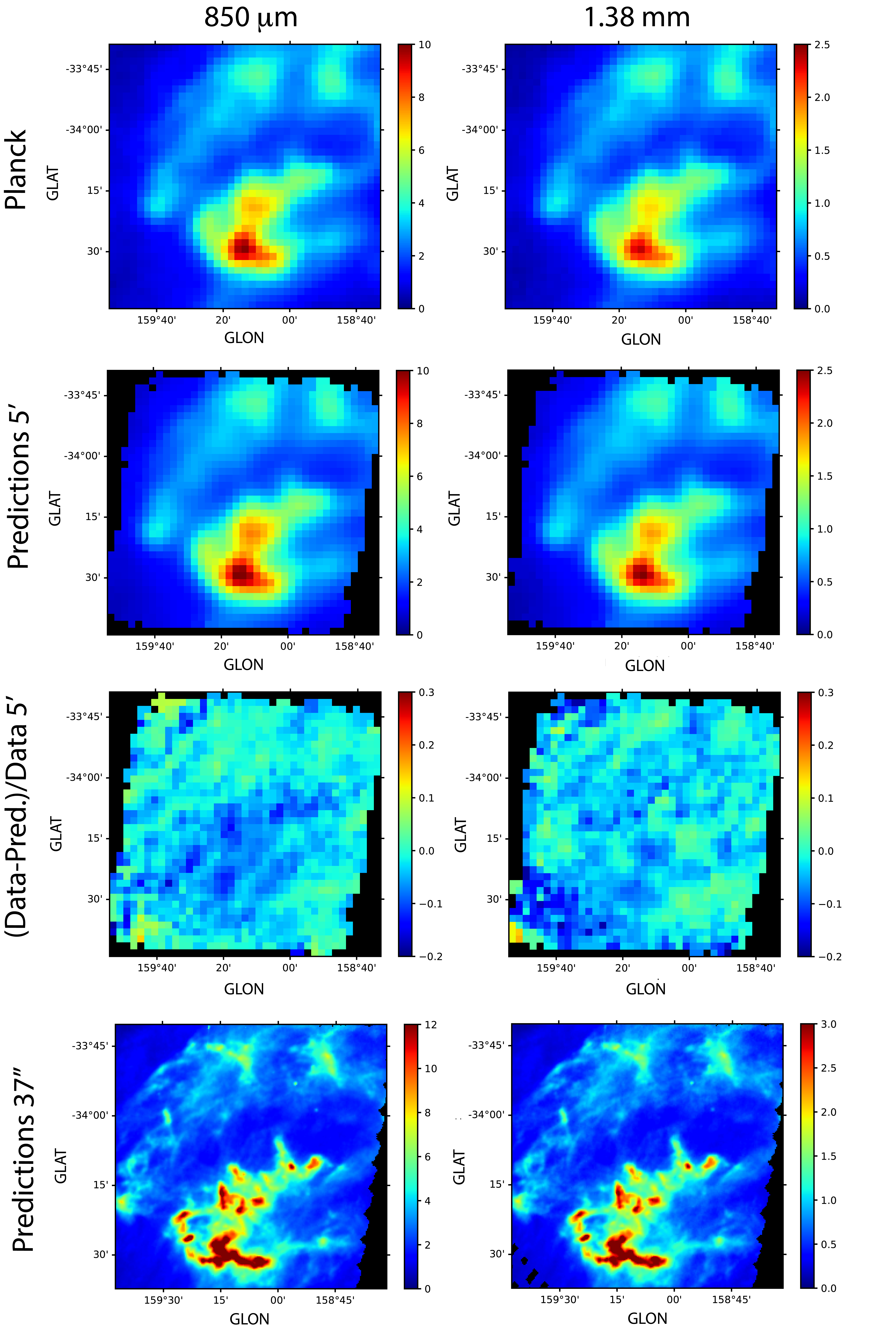}
\caption{Comparison between {\it Planck} data
  and predictions for MBM12, at 850
  $\mic$ on the left and 1.38 mm on the right. From top to bottom: {\it Planck} data (5$^{\prime}$ angular
  resolution), neural network predictions
  (5$^{\prime}$), relative error between data and predictions
  (5$^{\prime}$) and neural network predictions (37$^{\prime\prime}$). \label{fig_mbm12} }
\end{center}
\end{figure}

Prediction maps were generated in the HEALPix format with NSIDE=2048. We extracted some
selected regions for various astronomical features (filaments, cold cores, and star-forming regions)
or galaxies using the DrizzWeb interface (see Sect. 5.1). We compare predictions
and observations in the Galactic plane (Fig. \ref{fig_gal}), LMC (Fig. \ref{fig_lmc}),
SMC (Fig. \ref{fig_smc}), M31 (Fig. \ref{fig_m31}), M33
(Fig. \ref{fig_m33}), Orion (Fig. \ref{fig_orion}), Taurus
(Fig. \ref{fig_taurus}), LDN1642 (Fig. \ref{fig_ldn1642}), L134
(Fig. \ref{fig_l134}), L183 (Fig. \ref{fig_l183}), and MBM12
(Fig. \ref{fig_mbm12}). The relative errors between the data and the
predictions are also shown in each figure. Pixels in black correspond
to pixels that have been removed, either because of negative values in
the inputs maps, or because of errors greater than 20$\%$
(mainly in very bright regions or, conversely, in faint noisy
regions).

\subsection{Prediction maps at 37$^{\prime \prime}$}
 The main goal of this work is to provide prediction maps at higher
 resolution (37 $^{\prime \prime}$ here) in order to analyze large regions
of the sky
in the submm/mm wavelength range, at the {\it Herschel} resolution instead
of the {\it Planck} one.
In addition, prediction maps at 37$^{\prime \prime}$
are crucial to selecting potential regions for proposed observations with ground-based telescopes at higher resolution.
The regions used to train the neural network mix different environments, with
different distances and obviously different spatial scales. We
  do not expect to observe significant changes in the SED
  from an angular resolution of 5$^{\prime}$ to 37$^{\prime \prime}$ 
  for most regions of the Galaxy, or any significant 
  impact from free-free emission at 1.38 mm at 37$^{\prime \prime}$ resolution, except possibly in a few very bright regions. This is particularly true in the Orion region, where dust temperatures can reach as high as 70 K \citep{Arab12}. However, these very bright pixels in Orion (and in a few other UCHII regions) are masked in the final maps due to their relative errors exceeding 20$\%$ (see Sect. \ref{sec_pred_5arcm}), regardless of any potential contamination. A strong contribution of free-free emission could occur at angular resolutions of a few arcseconds.

We
therefore consider that the regions we used from training the neural network reveal
a large diversity of dust emission spectra that could be representative
of the overall of dust emission properties at long wavelengths. 
In consequence, we applied the parameters of the neural network that
give the best agreement with the {\it Planck} data to the 37$^{\prime
  \prime}$ {\it Herschel} input maps. These new prediction maps (with NSIDE=16384) are presented in all previous
figures (from \ref{fig_lmc} to \ref{fig_mbm12}). This way, we can clearly see the benefit of a
better angular resolution, compared to the {\it Planck} one.  

We next compare the prediction maps with available ground-based data, from the
JCMT/SCUBA-2 850 $\mic$ and Bolocam data at 1.1 mm. Because of the
filtering, we compare brightnesses in aperture photometry after removing the same
background in each maps.

\subsubsection{Comparison of cold core predictions with JCMT/SCUBA-2}
\begin{figure}
\begin{center}
\includegraphics[width=9cm]{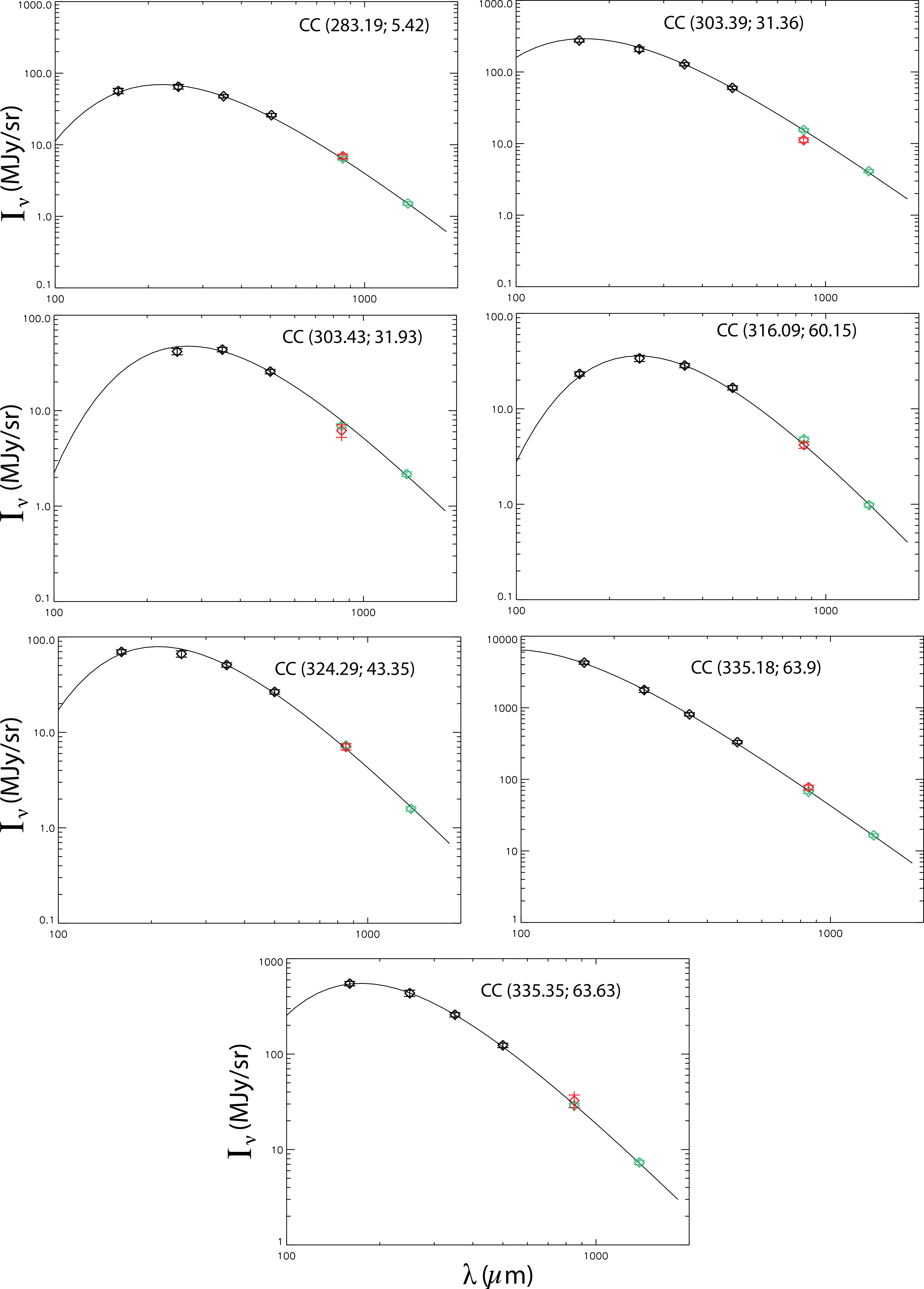}
\caption{SEDs of Cold Cores (Galactic coordinates given in the
  parenthesis), with {\it Herschel} data (from 160 to 500 $\mic$) in black,
  JCMT (850 $\mic$) in red, and predictions (850 and 1380 $\mic$) in green. A modified black-body fit is shown for convenience.\label{spec_jcmt} }
\end{center}
\end{figure}

Figure \ref{spec_jcmt} presents the extracted SED for each source (see
Sect. \ref{sec_sed_extract}). An additional modified black-body fit was adjusted to
give a simple view of the spectral shape of the SEDs. The difference between
JCMT 850 $\mic$ data and the predictions is less than 10$\%$ for
most of the source; therefore, the data and predictions are compatible within the
error bars, except for one source that evidences a difference of
less than 30$\%$.The comparison is nevertheless impressive, with the
37$^{\prime \prime}$ predictions and JCMT data showing similar
brightnesses. This agreement suggests that the dust emission
  behavior observed at an angular resolution of 5$^{\prime}$ across a wide
  range of pixels with various physical conditions can be reliably
  reproduced at sensibly higher resolution. However, predictions at 850 $\mic$
  do not systematically follow the modified black-body emission model. For instance,
  one source (RA=303.43; Dec=31.93) shows a 850 prediction that falls below the fit, while predictions for
  two other sources
  (RA=316.09; Dec=60.15 and RA=324.29; Dec=43.35) appear
  to slightly exceed the fit. 
The low JCMT brightness of one source is probably due to
an inaccurate flux calibration. For instance, the 450 $\mic$ cannot be used due
to its low accuracy in the flux calibration. 
\subsubsection{Comparison of UCHII regions predictions with Bolocam}
\begin{figure*}
\begin{center}
\includegraphics[width=18cm]{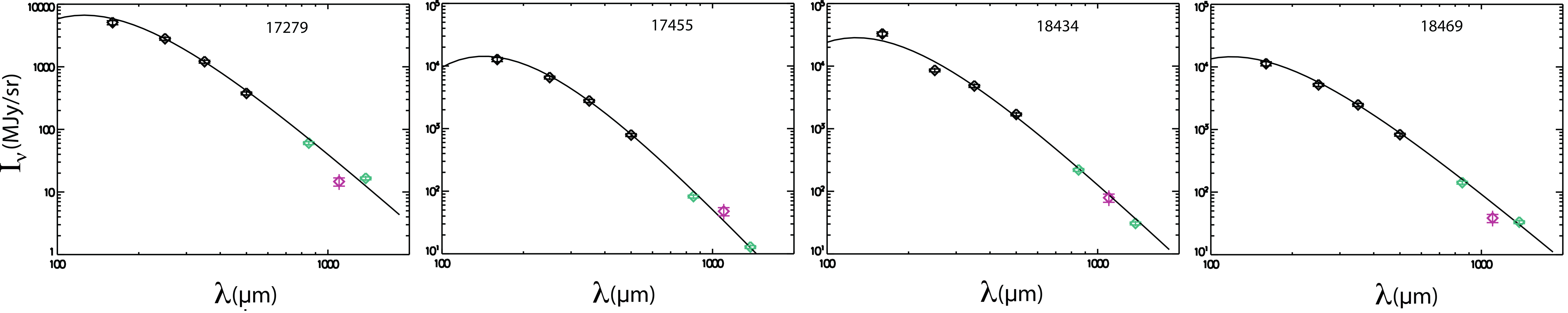}
\caption{SEDs of four UCHII regions \citep[coordinates given
  in][]{Paradis14}, with {\it Herschel} data (from 160 to 500 $\mic$) in
  black, Bolocam (1.1 mm) in pink, and predictions (850 and 1380 $\mic$)
  in green. A modified black-body fit is shown for convenience. See
  Fig. \ref{spec_bolocam_uchii} for the SEDs of the 12 UCHII regions
  described in Sect. \ref{sec_bolocam} and in \citet{Paradis14}. \label{spec_bolocam_uchii_cut} }
\end{center}
\end{figure*}

To compare the 1.38 mm prediction maps with observational data, we used
Bolocam data at 1.1 mm. However, these data suffer from an important
attenuation of the extended emission. For that reason, data from UCHII
regions (except perhaps from the Galactic center, Orion bar, and 30-Doradus) should be more reliable, since they are compact and their
emission largely dominates any potential contribution from the extended emission. The
comparison is given in Fig. \ref{spec_bolocam_uchii_cut} for four UCHII
regions, and in Fig. \ref{spec_bolocam_uchii} for the 12 UCHII
regions of this study. Except for one
UCHII region, the Bolocam data are most of the time slightly below
the modified black-body fit. Only three SEDs of UCHII show a significant difference
between Bolocam and  the fits (17279, 18469, and 18502). For the rest
of the UCHII regions, the comparison is quite favorable and strongly
suggests a compatibility between the predictions and Bolocam data. And
again, these results confirm the reliability of the prediction maps.  
Furthermore, the prediction maps do not suffer from filtering, or from other
data processing artifacts, and therefore can bring a lot advantages and
reliability when compared to ground-based data. 

\subsubsection{Prediction map delivery}
The prediction maps generated at angular resolutions of 5$^{\prime}$ and 37$^{\prime \prime}$
are provided on the CADE
website,\footnote{https://cade.irap.omp.eu/dokuwiki/doku.php?id=nnpredictions}
as well as the {\it Planck} data at 850 $\mic$ and 1.38 mm after CMB, CIB,
and CO removal (see Sect. \ref{sec_planck_data}). 

\section{Discussion}
\label{sec_disc}
\subsection{From the I$_{\nu}$(850)/I$_{\nu}$(1380) ratio to the spectral index and dust emissivity index}
\begin{figure*}
\sidecaption
\includegraphics[width=12cm]{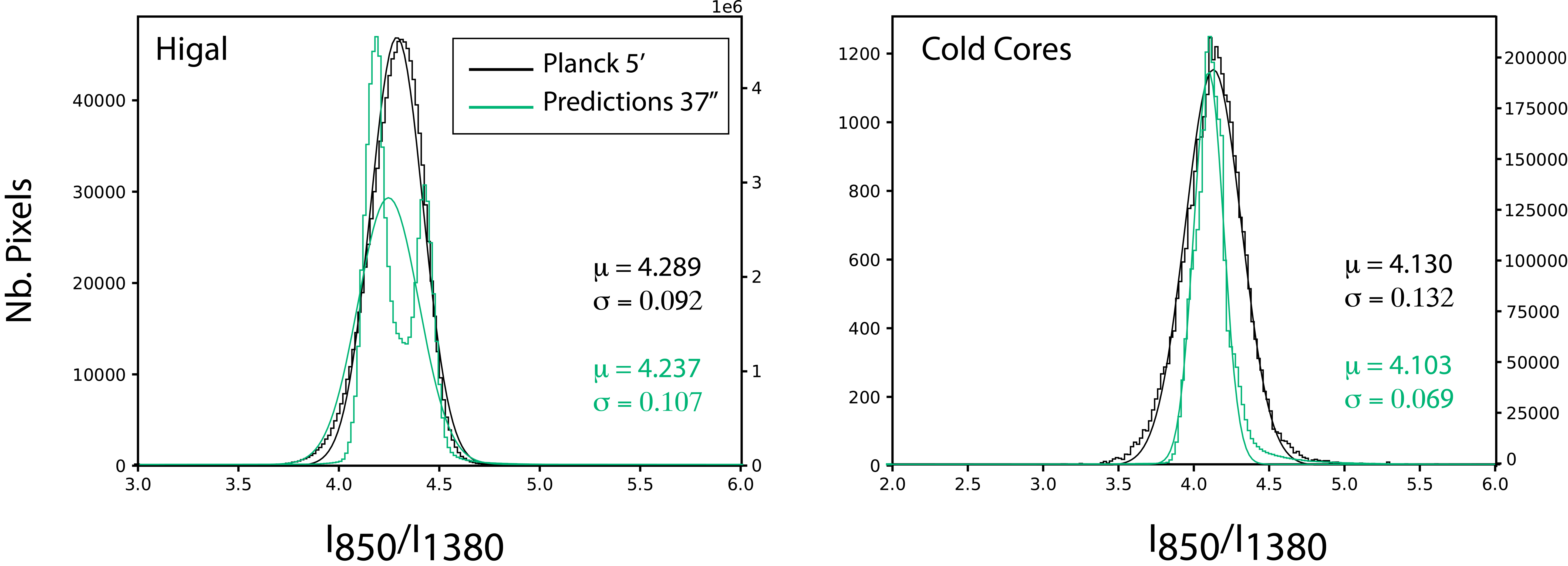}
\caption{Histograms of the I$_{\nu}$(850)/I$_{\nu}$(1380) ratio for
  the two Hi-GAL and Cold Cores
  {\it Herschel} large programs, deduced from
  the {\it Planck} data at 5$^{\prime}$ in black, and from the predictions at 37$^{\prime \prime}$ in
  green. Gaussian fits are overplotted, with the central values
  ($\mu$) and the standard deviations ($\sigma$) given in each
  panel. The left and right y axes in each panel correspond to the
  number of pixels for the {\it Planck} data and prediction
  histograms. See Fig. \ref{hist_comp} for histograms of all the {\it
    Herschel} large programs described in Sect. \ref{sec_herschel}.\label{hist_comp_cut} }
\end{figure*}

We first computed the I$_{\nu}$(850)/I$_{\nu}$(1380) ratios derived from {\it Planck} data
for each survey, and derived from the prediction maps at
37$^{\prime\prime}$. The histograms and Gaussian fits of the Hi-GAL
and Cold Cores programs are shown in Fig. \ref{hist_comp_cut}, and
of all the programs described in Sect. \ref{sec_herschel}
in Figure \ref{hist_comp}. The first difference appears for the Hi-GAL and Gould
Belt surveys, which evidence distinct histograms: a Gaussian profile
in the case of {\it Planck} data, and a double Gaussian profile in the case
of predictions. This difference does not come from the resolution,
because we observe the same histograms when using predictions at
5$^{\prime}$ (not shown in the paper), but rather from an
  intrinsic effect of the neural network. The neural network slightly
  modified the peak of the I$_{\nu}$(850)/I$_{\nu}$(1380) histogram ratio. This effect is the result of small variations of
a few percent in the predictions compared to the data. The two peaks
in the prediction histogram correspond to a decrease or increase of a few percent
($\sim$3-4$\%$ compared to the single peak centered at 4.29 in the
{\it Planck} histogram of Hi-GAL at 5$^{\prime}$, for instance) induced by the neural network,
which is negligible. In the other cases, the 37$^{\prime\prime}$ prediction
histograms show narrower profiles, except for the VNGS survey,
  which exhibits the opposite behavior, likely due to there being significantly
fewer pixels in the histograms. These differences
  in histogram profiles result from an inherent effect of the neural
network. The extreme dispersion of the ratio is approximately 13$\%$
at the full width at half maximum (in the VNGS histograms). However, the most important result here is that the central values
of the histograms are quite similar for all environments (see the
$\mu$ values in the figure). This means that statistically the ratio
is between 4 and 4.3. From a general point of view, the mean slope in the
submm-mm wavelength range is statistically the same whatever the
environment, indicating a certain stability in the dust emission spectrum
at long wavelengths. We note however that the most diffuse part of the
Galactic ISM, used to derive the ``standard'' Galactic
SED, is not represented in this analysis.  

We computed the observed spectral index, $\alpha_{850-1380\mic, obs}$, for
all {\it Herschel} large programs using {\it Planck} data at 5$^{\prime}$, defined as
\begin{equation}
  \alpha_{850-1380\mic} = \frac{log \left( I_{\nu}(850)/I_{\nu}(1380)
    \right) }{log(850/1380)}
  .\end{equation}
Assuming optically thin emission at long wavelengths and in the
Rayleigh-Jeans approximation, we can deduce the dust emissivity index ($\beta^{\star}$)
by using this simple relation: $\beta^{\star}$=$\alpha$-2. For
simplicity, we note $\beta^{\star}$, the dust emissivity index derived
from the I$_{\nu}$(850)/I$_{\nu}$(1380) ratio, compared to the standard $\beta$
derived from a temperature (T)-spectral index ($\beta$) model. We obtain
$\beta_{850-1380\mic}^{\star}$ values around 1, which indicates a low
emissivity index at long wavelengths. Tab \ref{table_corr} gives the
correspondence between I$_{\nu}$(850)/I$_{\nu}$(1380) 
and $\beta_{850-1380\mic}^{\star}$ for the central value $(\mu)$ of each
histogram. However, we note that computing
  $\beta_{850-1380\mic}^{\star}$ using predictions (at
  37$^{\prime\prime}$ or 5$^{\prime}$) can be biased by the neural
  network due to variations in the
  histogram profiles (see above). These variations could lead to a 
  change of spectral index up to $+0.25$ in extreme cases, observed in the VNGS survey. 
\begin{table}
  \caption{Correspondence between the intensity ratio derived from
    {\it Planck} (5$^{\prime}$) and the
    dust emissivity index.\label{table_corr}}
  \begin{center}
    \begin{tabular}{lcc}
\hline
      \hline
      Region  &  I(850)/I(1380) &
                                                       $\beta_{850-1380\mic}^{\star}$ \\
    & {\it Planck} (5$^{\prime}$)& \\  
      \hline
      Hi-GAL & 4.289& 1.005 \\
      Cold Cores & 4.130& 0.927 \\
      Gould Belt & 4.110 & 0.917 \\
      Heritage & 4.198& 0.960 \\
      HerM33es & 4.202& 0.962 \\
      Helga & 4.247& 0.984 \\
      KINGFISH &4.122 &0.922 \\
      VNGS & 3.960& 0.840 \\
       \hline
      \end{tabular}
  \end{center}
 \end{table} 
Statistically, several analyses evidenced
significantly higher $\beta$ values at shorter wavelengths
\citep[see for instance][]{Stepnik03, Abergel10, Paradis10, Juvela11} — that is, in the FIR wavelength range — meaning
that a break in the spectral shape in the dust emission spectrum is a
realistic assumption, as has already been proposed by several authors
\citep[see for instance][]{Meny07, Paradis09, Juvela15}. \\
\citet{Reach95} tried to explain 
  the flattening in the COBE/FIRAS spectrum \citep{Fixsen94} of the
  diffuse ISM. They found a good fit to 
the data, with the use of a two modified black-body model
  with dust populations at $\sim 16-21$ K and $\sim 4-7$ K. The
    authors, however, argued against the existence of such a cold dust
    component. \citet{Finkbeiner99} also provided a two-component dust model, in which the
  components are described by two distinct emissivity spectral
  indices. This model required amorphous silicate grains at 9.5 K ($\beta=1.7$) and
  carbonaceous grains at around 16 K ($\beta=2.7$). The {\it Planck}
  2014 data release was analyzed by \citet{Meisner15}, who proposed
  an update of the \citet{Finkbeiner99} model, with best temperatures
  of 15.7 K ($\beta$=2.82) and 9.75 K ($\beta$=1.63), as a better
  model than the \citet{PlanckXI} model (T=19.74 K,
  $\beta$=1.6). 

Adding long-wavelength data (in the mm range) usually decreases the
global $\beta$
value, mainly in the diffuse medium \citep[see for instance][]{PlanckXVII}). However, $\beta$ values have always been derived
including FIR wavelengths, which induces an increase in $\beta$ compared to the use of the submm-mm range only. This work
  reinforces the idea that the dust emissivity spectral
  index could be wavelength-dependent, as has been observed in laboratory
  experiments \citep[see for instance][]{Agladze94, Agladze96,
    Mennella98, Boudet05, Coupeaud11, Demyk17a, Demyk17b}. However, thermal free-free emission of ionized gas
  could also be responsible for this behavior by contaminating the submm-mm
  data. For instance, \citet{Izotov14}
  studied dust emission in a large sample of emission-line
  star-forming galaxies. They found important free-free emission of
  ionized gas in the submm and mm range that could cause the submm excess, as was
  discussed by \citet{Remy-Ruyer13}. \citet{Lisenfeld02} and
  \citet{Galliano05} also analyzed the contamination by free-free emission in dwarf galaxies and identified a contribution that can reach
  13$\%$ of the emission at 850 $\mic$ and 23$\%$ at 1.2 mm. However, in the Magellanic
Clouds, the different components such as free-free, synchrotron, and
foreground emissions do not really affect the 850 $\mic$ and
  1.38 emissions in the LMC and
  SMC SEDs after CMB subtraction \citep[see Table 2 in][]{Planck11LMC}. Their
  contribution may affect the SED at longer wavelengths. In addition,
  we do not expect to observe free-free nor synchrotron emission in cold cores. If the low emissivity index were the result of these
  additional components, we would therefore statistically expect to
  observe important variations from one environment to the other,
  which is not the case. 

 Moreover, $\beta_{850-1380\mic}^{\star}$ seems to be much lower than has previously been thought. For instance, \citet{PlanckXI} derived a mean spectral index of 1.62 over
  the whole sky using a modified black-body model between 100 and 850
  $\mic$. First, this value of
  $\beta$ is highly temperature-dependent with this model, as well as uncertainty-dependent, and was
  computed on a shorter wavelength range compared to our analysis,
  which could explain a higher value. Nevertheless, looking at their Fig. 9,
  we observe $\beta$ values going from $\sim$1.2 to 2.2 along the
  Galactic plane, and with high values in the inner Galactic
  plane. Our lower value could be recovered assuming a flattening of the dust
  emission spectrum for wavelengths larger than 850 $\mic$.
\citet{PlanckXVII} confirm this behavior by comparing
  the $\beta_{mm}$ ($\lambda$ $>$ 850 $\mic$) with the $\beta_{FIR}$.
  We note, however, that the $\beta_{850-1380\mic}^{\star}$ that we obtain is a strict value derived
  from a brightness ratio and does not take any uncertainty into
  account. The inclusion of FIR data that requires knowing the dust
  temperature would probably induce a higher
  $\beta$ value. If we compute $\beta_{850-1380\mic}^{\star}$ in the LMC and SMC
  derived from the global SED using {\it Planck} data from \citet{Planck11LMC} (see their table 2), we get values of
  1.02 and 0.67 in the LMC and SMC, respectively, significantly lower than the
  values derived from a T-$\beta$ model applied from FIR to mm
  wavelengths, which are 1.5 and 1.2,
  respectively. For the SMC, the authors observed a significant
  flattening of the dust emission for
  $\lambda$ $>$ 850 $\mic$ that could explain the lower value of
  $\beta_{850-1380\mic}^{\star}$ based 
  on the I$_{\nu}$(850)/I$_{\nu}$(1380) ratio. For the
  LMC, we note that $\beta$ derived from a T-$\beta$ fit is not well constrained when looking at the $\chi^2$
  values, and is highly degenerate with the dust temperature
  value. 

For M33, \citet{Tibbs18} recovered an effective dust emissivity index
derived from the 100 $\mic$ to 3 mm data
going from 0.93 to 1.44 depending on the CMB-subtraction method,
whereas, using their {\it Planck} brightness values, we computed $\beta_{850-1380\mic}^{\star}$ between
0.87 and 1.1. 
For the same galaxy, we find $\beta_{850-1380\mic}^{\star}$ of 0.98 using
values from \citet{Hermelo16} (see their Table 2).

We can conclude that the dust spectral index seems to vary with
wavelength, with a FIR dust emissivity index that could be significantly higher than
its value at long wavelengths. However, at a different spatial scale
and in particular at an angular resolution of a few arcseconds, we do not
know if this spectral behavior can still occur.
\subsection{Predicting dust emission in nearby extragalactic environments}
One of the most impressive results is that even if the neural network has been
trained on Galactic data only (Hi-GAL, Gould Belt, and Cold Cores), it
is able to predict dust emission in other nearby galaxies, and in the
Magellanic Clouds in particular (Fig. \ref{fig_lmc} and
\ref{fig_smc}). The Magellanic Clouds are two of the nearest
galaxies and as such are easily resolved by observations. These
galaxies are characterized by a lower metallicity compared to the
MW \citep[$\frac{1}{2}Z_{\sun}$ and $\frac{1}{5}Z_{\sun}$ for the
LMC and SMC,][]{Russel92} . The physical conditions in
the Magellanic Clouds are different from the MW: the radiation field
is higher \citep{Dufour84, Lequeux84}, the filling factor of dense clouds is lower in the
Magellanic clouds than in the MW \citep{Pineda12}, and the average dust
emission spectrum appears significantly flattened in the FIR-submm range in
the LMC and even more in the SMC 
\citep{Israel10, Bot10, Remy-Ruyer13}. Therefore, it is accepted that the dust properties differ from those of Galactic dust, though the origin of this difference remains unexplained. Each model that has its own dust properties (abundances,
sizes, optical properties, etc.) can always reproduce (more or less)
the NIR to submm/mm data by changing the
parameters \citep{Chastenet17, Paradis23}. It is possible to interpret the emissivity variations in
the framework of one model and reproduce them by changing the
parameters, but in the end we are not able to 
determine the dust emission brightness at a specific wavelength for a given
region of the sky. Conversely, the neural network seems to be
able to predict the dust emission in various environments by analyzing only
the dust spectral behavior between 160 $\mic$ and 500$\mic$. That
means that the submm/mm emission is fully predictable using the FIR
domain. Therefore, the submm excess already appears in the FIR
wavelength range. For instance, several analyses explored the 500
$\mic$ excess \citep{Paradis12b, Gordon14, Remy-Ruyer13}, which is also a crucial wavelength at which a change
in the emissivity spectral index with wavelengths was reported in the
framework of dust modeling \citep{Meny07}.
\begin{figure}
\begin{center}
\includegraphics[width=8cm]{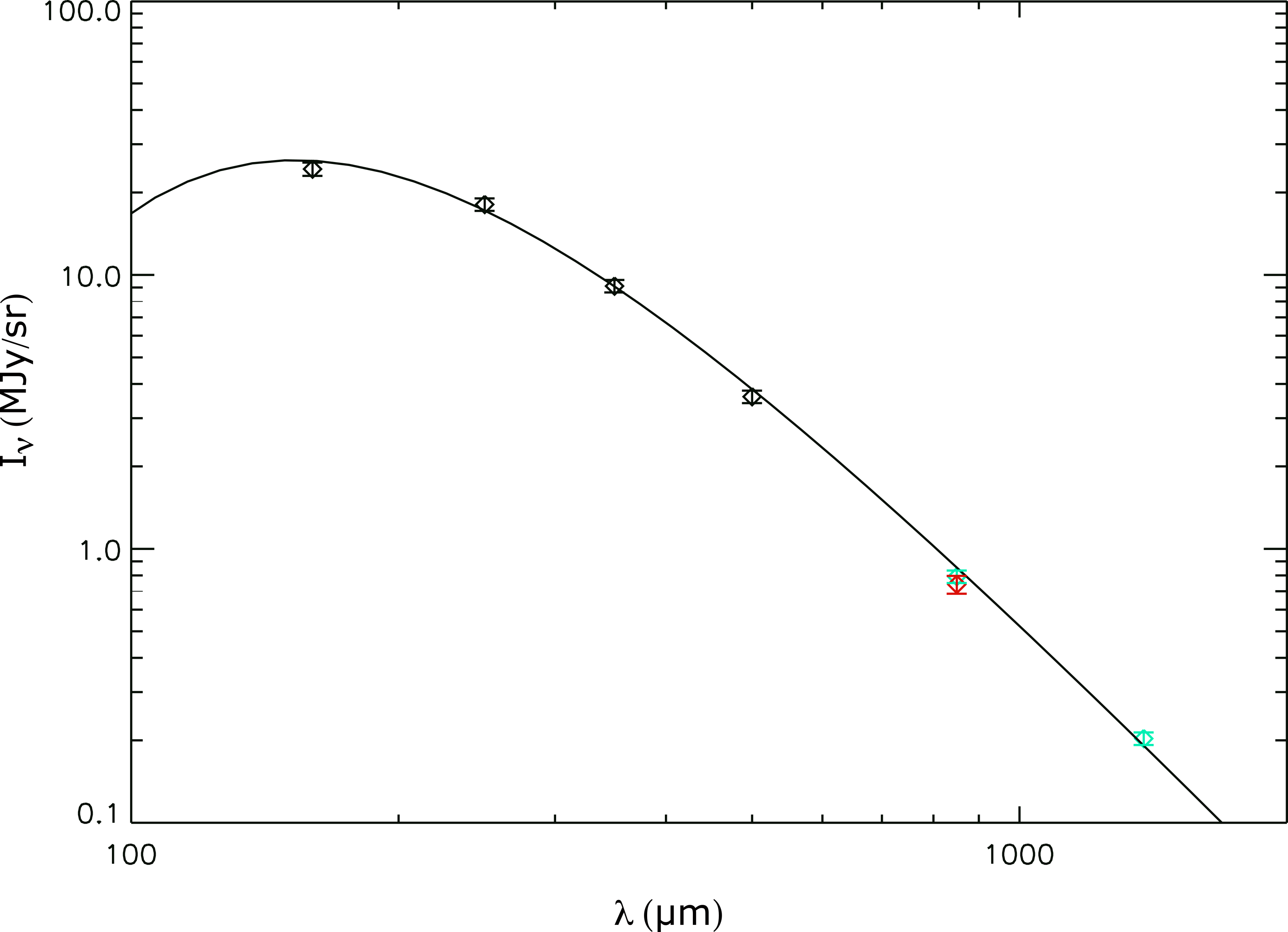}
\caption{SED of M31, with {\it Herschel} data (from 160 to 500 $\mic$) in
  black, JCMT (850 $\mic$) in red, and predictions (850 and 1380 $\mic$) in green. A modified black-body fit is shown for convenience.\label{sed_m31} }
\end{center}
\end{figure}

In a similar way, predictions at 5$^{\prime}$ are very convincing for
M31 and M33 (Fig. \ref{fig_m31} and \ref{fig_m33}). Several analyses
of dust emission in M31 and M33 did not converge on the variation in the
dust opacity index in dense environments using {\it Herschel}
images. 
\citet{Tabatabaei14} showed a decrease in the spectral index with
galactocentric distance in M33, whereas \citet{Smith12} evidenced the
opposite behavior as a function of radius in M31, both with {\it Herschel}
data and a similar linear resolution. This discrepancy could be the 
result of the degeneracy between the dust temperature and the spectral
index in M31, as was pointed out by \citet{Tabatabaei14}. This
point shows the complexity of understanding the dust emission
behavior, and so of determining the dust temperature and dust
masses. With the lack of long-wavelength data at the {\it Herschel}
resolution, no one is capable of predicting the spectral behavior
of the dust emission in the mm range. 

The SED of M31 combining {\it Herschel}, SCUBA-2, and predictions is given in
Fig. \ref{sed_m31}. We can note that predictions at 850 $\mic$ are in
very good agreement with SCUBA-2 data. This confirms again the goodness
of the calculated prediction maps. Moreover, these maps do not suffer from extended
emission filtering; the noise level is also significantly reduced
compared to ground-based data. Prediction maps of the KINGFISH survey also evidence good agreement when
comparing with {\it Planck} data in Figs. \ref{corr_850_app} and \ref{corr_1p4_app}. Comparisons
with the VNGS predictions show, however, some discrepancies for a few
pixels that have then been removed in the final maps produced.

If the neural network is able to predict emission in any region of the sky,
this means that the spectral behaviors of the dust emission in the
FIR-mm are not necessarily too different from one
nearby galaxy to another. Whatever the composition of the large dust
grains and their dust properties, their spectral behaviors are included in
the FIR-mm global emission and can be reproduced across various
environments.
Or, within a simple hypothesis, the large dust components could
have similar dust properties in our Galaxy and in some nearby
extragalactic environments, but their proportions could spatially vary.

\section{Conclusions}
\label{sec_cl}
By applying neural networks to large datasets from the {\it Herschel} and {\it Planck}
missions, we have produced, for the first time, prediction maps of dust
emission at a 37$^{\prime \prime}$ angular resolution, in the
two {\it Planck} bands centered at 850 $\mic$ (353 GHz)
and 1.38 mm (217 GHz). We are making these maps available to the community through the CADE service, covering surveys such as Hi-GAL, Cold Cores, Gould Belt, Magellanic Clouds, M33, M31, KINGFISH, and VNGS. We estimate
the uncertainties on these predictions to be approximately 4$\%$ at 850
$\mic$ and 7$\%$ at 1.38 mm. 
Although the supervised deep learning algorithms were trained primarily on Galactic environments, the neural networks are also capable of accurately reproducing data from nearby extragalactic environments. This impressive result
suggests that variations in dust properties could be reproducible across
different regions of the sky, provided that appropriate training datasets are used. In that sense, the spectral behaviors of the
FIR to the mm dust emission are similar
in Galactic and nearby extragalactic environments.

The prediction maps have been compared with JCMT/SCUBA-2 observations
at 850 $\mic$, as well as with Bolocam data at 1.1 mm. The agreement
between predictions and SCUBA-2 data is very good (less than 10$\%$ for
most of the sources and within the error bars). Bolocam data are also
compatible with predictions of UCHIIs. Another important result is the low spectral index between 850 $\mic$
and 1.38 mm, statistically close to 1, which could support the
hypothesis that dust emission flattens at long wavelengths.

The large
spatial coverage of the prediction maps will help in statistically probing the
dust emission 
variations at long wavelengths in specific regions due to the important benefit for the
angular resolution compared to {\it Planck}. In addition, these maps
could play the role of first templates of dust emission at this resolution in the framework of
foreground subtraction. Moreover, they
will help in preparing future observations at high angular resolutions.
The powerful capabilities of neural networks could allow future
analyses to make predictions in different regions of the sky, and/or at
other wavelengths. As a second step, it would be promising to apply
deep learning techniques, including convolutional neural networks, to recover the best
angular resolution of the data used in the process.   

\begin{acknowledgements}
We acknowledge the referee for his constructive and valuable comments. 
We acknowledge the use of data and analysis softwares provided by the
Centre d'Analyse de Données Etendues (CADE), a service of
IRAP-UPS/CNRS (http://cade.irap.omp.eu). DP acknowledges Emmanuel
Caux and Philippe Garnier for fruitful discussions on Machine
Learning.
\end{acknowledgements}

\begin{appendix}
  \section{Additional material}
\begin{figure}[!h]
\begin{center}
  \includegraphics[width=9cm]{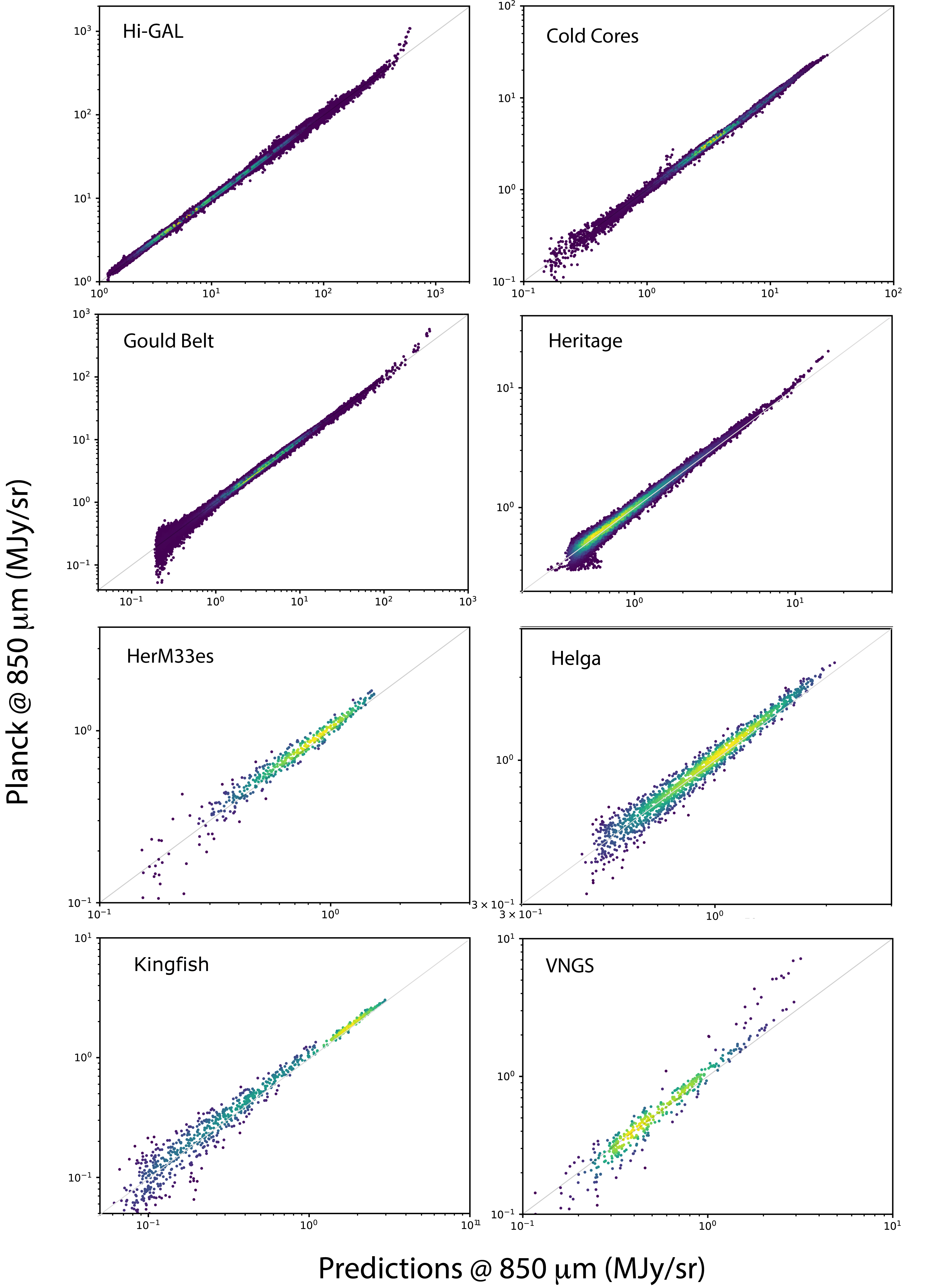}
\caption{Correlation plots between {\it Planck} data and neural network predictions at
  850 $\mic$, for the different {\it Herschel} large programs.  \label{corr_850_app} }
\end{center}
\end{figure}
\begin{figure}
\begin{center}
\includegraphics[width=9cm]{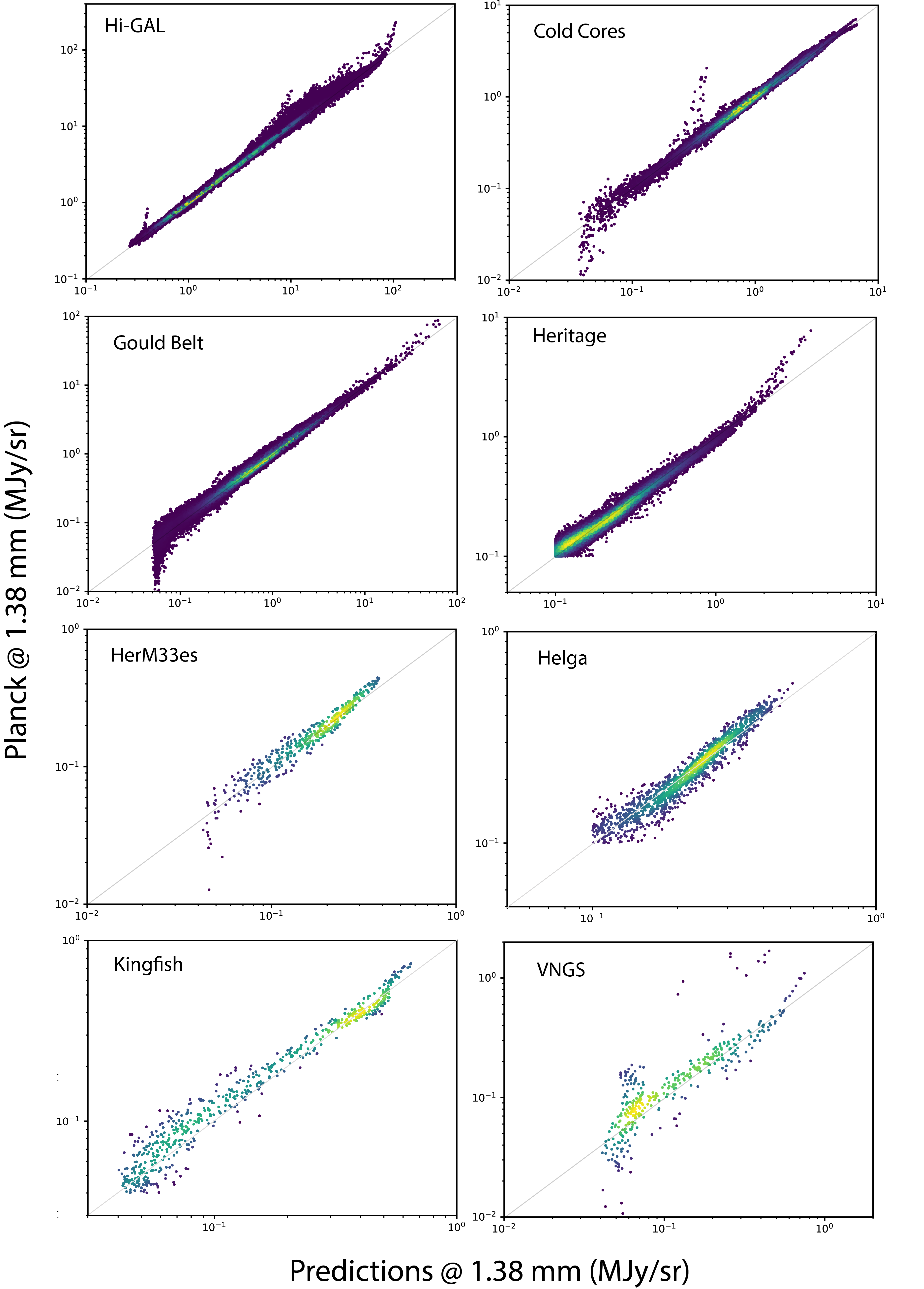}
\caption{Correlation plots between {\it Planck} data and neural network predictions at
  1.38 mm, for the different {\it Herschel} large programs. \label{corr_1p4_app} }
\end{center}
\end{figure}

\begin{figure*}[!h]
\begin{center}
\includegraphics[width=18cm]{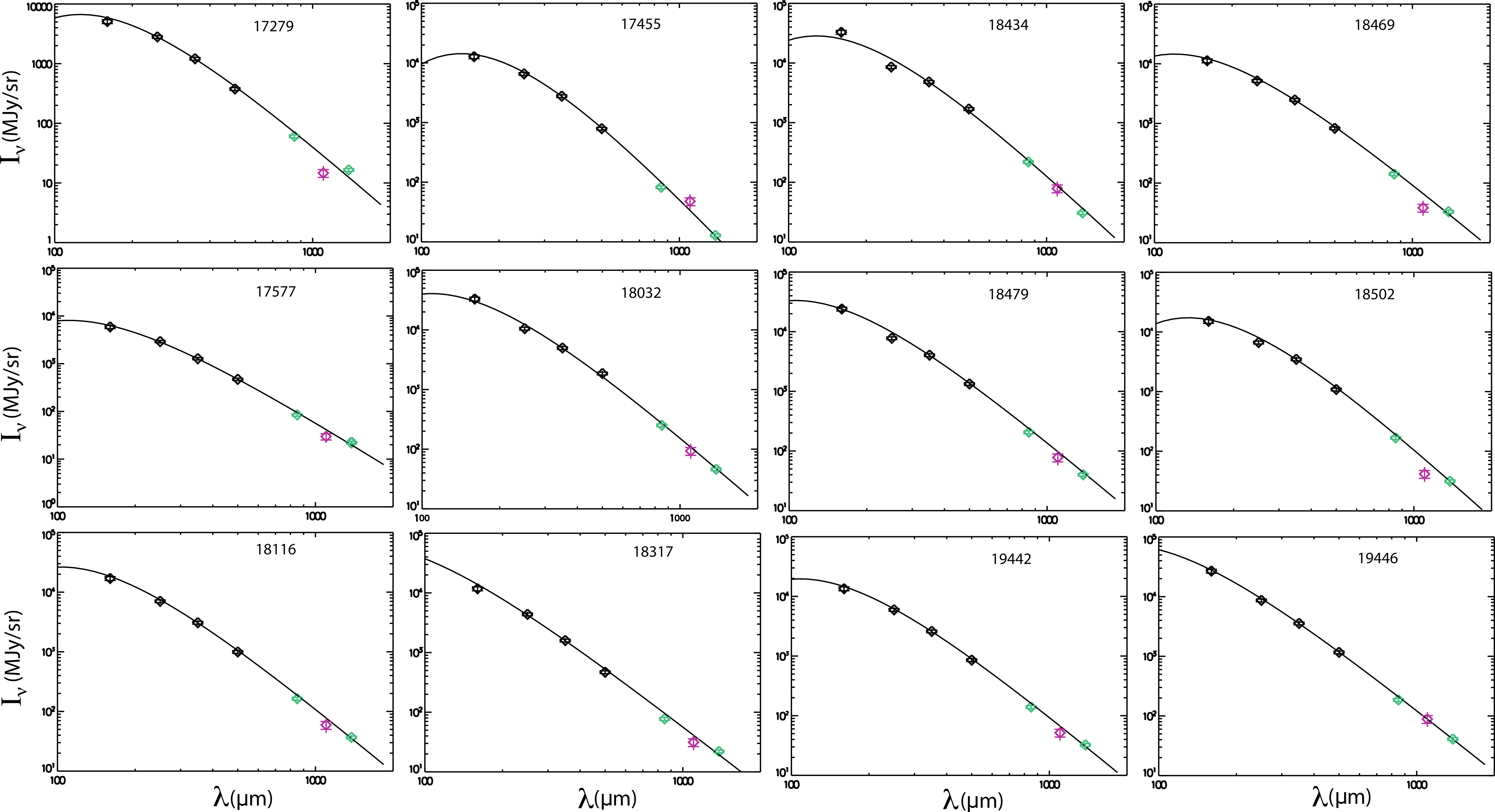}
\caption{SEDs of twelve UCHII regions \citep[coordinates given
  in][]{Paradis14}, with {\it Herschel} data (from 160 to 500 $\mic$) in
  black, Bolocam (1.1 mm) in pink, and predictions (850 and 1380 $\mic$) in green. A modified black-body fit is shown for convenience.\label{spec_bolocam_uchii} }
\end{center}
\end{figure*}

\begin{figure*}
\sidecaption
\includegraphics[width=12cm]{Hist_850o1p4_comp.png}
\caption{Histograms of the I$_{\nu}$(850)/I$_{\nu}$(1380) ratio for each
  {\it Herschel} large program, deduced from
  the {\it Planck} data at 5$^{\prime}$ in black, and from the predictions at 37$^{\prime \prime}$ in
  green. Gaussian fits are overplotted, with the central values
  ($\mu$) and the standard deviations ($\sigma$) given in each
  panel. The left and right y axes in each panel correspond to the
  number of pixels for the {\it Planck} data and prediction
  histograms.\label{hist_comp} }
\end{figure*}

\end{appendix}

\end{document}